\documentclass[notitlepage,superscriptaddress,aps,prd,nobibnotes,nofootinbib,10pt,groupedaddress,longbibliography]{revtex4-1}
\usepackage{braket,verbatim,bm,bbold,amsmath,amssymb,epsfig,xcolor}
\definecolor{redCB}{RGB}{150,50,60}  
\definecolor{dteal}{RGB}{7,94,84}  
\definecolor{blueCB}{RGB}{108,144,170}
\usepackage[colorlinks=true,    	
			pdfstartview=FitV,
            bookmarks=false,
            citecolor=dteal,
            linkcolor=dteal,
           linkcolor=dteal,
            hyperfootnotes=true,
            linktoc=page,
            urlcolor=dteal]{hyperref}
\renewcommand{\footnoterule}{%
  \kern-6pt
  \hrule width .5in height .4pt
  \kern6pt
}
\usepackage{xparse}
\usepackage{soul}
\usepackage{tikzlings}
\usepackage{tikzducks}
\usepackage{flushend}
\usepackage{nicematrix}
\usepackage{chngcntr}
\usepackage{multirow,appendix,mathtools}
\newcommand\scalemath[2]{\scalebox{#1}{\mbox{\ensuremath{\displaystyle #2}}}}
\newcommand{\nc}{\newcommand}

\def\[{\left[}
\def\]{\right]}

\def\a{\alpha}

\def\D{\mathcal{D}}

\nc{\ir}{\mathrm{i}}
\nc{\dd}{\mathrm{d}} 
\nc{\eE}{\mathrm{e}}
\nc{\tr}{\text{Tr}}
\nc{\Tr}{\text{Tr}}
\nc{\id}{\mathbb{I}}
\nc{\Z}{\mathcal{Z}}
\nc{\E}{\mathcal{E}}
\nc{\F}{\mathcal{F}}
\nc{\Om}{\Omega}
\nc{\N}{\mathcal{N}}
\nc{\spp}{\hspace{1pt}}
\nc{\spm}{\hspace{-1pt}}
\nc{\lb}{\label} 
\nc{\nn}{\nonumber} 
\nc{\ra}{\rangle} 
\nc{\la}{\langle} 
\nc{\Nn}{\mathcal{N}} 
\nc{\sigb}{\boldsymbol{\sigma_{\text{bd}}}}
\nc{\Hh}{\mathcal{H}}
\nc{\HA}{\mathcal{H}_A}
\nc{\HB}{\mathcal{H}_B}
\nc{\HC}{\mathcal{H}_C}
\nc{\ta}{\tilde{a}}
\nc{\eps}{\varepsilon}
\nc{\Q}{\mathcal{Q}}
\nc{\C}{\mathcal{C}}
\nc{\T}{\mathcal{T}}
\nc{\Oo}{\mathcal{O}}
\nc{\bG}{\mathbf{G}}

\newcommand{\ii}{\mathrm{i}}

\newcommand{\cP}{\mathcal{P}}
\nc{\rank}{\text{rank}}



\def\bea#1\eea{\begin{align}#1\end{align}}
\def\bes#1\ees{\begin{equation}\begin{split}#1\end{split}\end{equation}}

\makeatletter
\renewcommand{\frontmatter@footnote@produce}{%
  \ifx\@FMN@list\@empty\else
    \begingroup
      \def\@thefnmark{}%
      \long\def\frontmatter@makefntext##1{\noindent##1\par}%
\let\email@sep\@empty
\def\@TBN@opr##1##2{%
  \email@sep
  \def\email@sep{,\quad}%
  \begingroup
    \csname c@\@mpfn\endcsname=##1\relax
    \def\@thefnmark{\frontmatter@thefootnote}%
    \Hy@raisedlink{%
      \hyper@anchorstart{frontmatter.##1}%
      \hyper@anchorend
    }%
    \@makefnmark\,##2%
  \endgroup
}
      \frontmatter@footnotetext{\@FMN@list\unskip}%
    \endgroup
    \global\let\@FMN@list\@empty
  \fi
}
\makeatother

\usetikzlibrary{matrix,backgrounds,fit}
\pgfdeclarelayer{myback}
\pgfsetlayers{myback,background,main}

\tikzset{mycolor/.style = {line width=1bp,color=#1}}%
\tikzset{myfillcolor/.style = {draw,fill=#1}}%

\NewDocumentCommand{\highlight}{O{blue!40} m m}{%
\draw[mycolor=#1] (#2.north west)rectangle (#3.south east);
}

\NewDocumentCommand{\fhighlight}{O{blue!40} m m}{%
\draw[myfillcolor=#1] (#2.north west)rectangle (#3.south east);
}

\begin{document}

\title{Multipartite entanglement structure of crystal quenches}

\author{Paul Gaudin}
\email{paul.gaudin@irsamc.ups-tlse.fr}
\affiliation{\vspace{3pt}Laboratoire de Physique Théorique, CNRS, Université de Toulouse, France}

\author{Cl\'ement Berthière}
\email{clement.berthiere@irsamc.ups-tlse.fr}
\affiliation{\vspace{3pt}Laboratoire de Physique Théorique, CNRS, Université de Toulouse, France}

\date{\today}

\begin{abstract}\vspace{4pt}
Understanding how entanglement emerges and spreads in out-of-equilibrium quantum many-body systems is a fundamental problem. In this paper, we investigate multipartite entanglement after quantum quenches. 
In particular, we study the entanglement structure and dynamics of quasiparticle multiplets generated by crystal quenches in the tight-binding chain. We identify zero-configurations for which a multiplet shared between three parties has no irreducible tripartite entanglement---these configurations consist entirely of pairwise distributed bipartite-entangled states. The existence of zero-configurations has important implications on the entanglement structure and dynamics of crystal quenches. For example, we show that it can further delay the production of tripartite entanglement. We build the quasiparticle picture for the Markov gap, and prove that it is a faithful witness of irreducible tripartite entanglement for crystal quenches. We discuss several key aspects of their tripartite entanglement dynamics through the lens of the Markov gap. We confront our results with exact lattice numerics and find perfect agreement in the hydrodynamic regime.
\end{abstract}

\maketitle

\makeatletter
\def\l@subsubsection#1#2{}
\makeatother

\reversemarginpar 


\section{Introduction}

Understanding the nonequilibrium dynamics of isolated quantum many-body systems is a central problem in condensed matter physics, statistical mechanics, and quantum information theory. A fundamental question is how entanglement and correlations develop and spread in such systems. A widely studied setting is the quantum quench \cite{1970PhRvA...2.1075B,DeChiara:2005wb,Calabrese:2007rg,2008JSMTE..05..018L,Polkovnikov:2010yn,Gogolin:2015gts,Calabrese:2016psf,Essler:2016ufo,Cotler:2016acd}, in which a system initially prepared out of equilibrium undergoes unitary evolution governed by a Hamiltonian.

In integrable models, this dynamics can be understood using hydrodynamic approaches.
Within the quasiparticle picture \cite{Calabrese:2005in,Alba:2017lvc,Alba:2017ekd,Calabrese_2020}, the initial state acts as a source of entangled quasiparticle excitations that propagate ballistically, spreading entanglement across the system. Much of the literature on quantum quenches has concentrated on bipartite entanglement with initial states producing pairs of quasiparticles \cite{Fagotti:2008mlc,Rieger:2011ipr,Blass:2012iun,Kim:2013etb,2014EL....10740002K,Alba:2017qag,Alba:2018hie,Bertini_2018,Chapman:2018hou,Ali:2018aon,Alba:2019ybw,Modak:2020faf,Parez:2021pgq,DiGiulio:2021oal,Lagnese:2021qqu,Parez:2022egh,Alba:2022enm,Santalla:2022ygq,Yamashika:2023qkr,Khetrapal:2024chv,Caceffo:2024qce,Chalas:2024yts,Travaglino:2024sxj,Maric:2024hzo}, and providing a simple explanation for the characteristic growth and saturation of entanglement entropy, as well as the onset time for mutual information and logarithmic negativity.  Recently, attention has shifted toward quenches producing entangled multiplets rather than pairs \cite{Bastianello:2018fvl,Bertini:2018ymp,Carollo:2022lrq,Berthiere:2024sio}, providing a natural setting to investigate multipartite entanglement.
The structure of entanglement becomes significantly more complex when more than two parties are involved \cite{Guhne:2008qic,Horodecki:2009zz}. For multiplets shared among three or more subsystems, determining the entanglement structure requires distinguishing inequivalent forms of multipartite entanglement. In particular, a pure state can be entangled across every bipartition while consisting entirely of bipartite-entangled states shared between pairs of subsystems. Understanding how these correlations emerge and spread is therefore essential for a more complete description of quantum many-body dynamics.

In this paper, we investigate multipartite entanglement after crystal quenches in the tight-binding chain. The initial crystal states consist of a repeated unit cell of size $p$ containing $q$ consecutive occupied sites and $p-q$ empty sites. In particular, we study the tripartite entanglement structure and dynamics of the quasiparticle multiplets generated by these initial pure states.
One of the main results of this work is the identification of special tripartitions of crystal multiplets, which we call \textit{zero-configurations}. These configurations exist when $p$ and $q$ are not coprime, and follow from the particular structure of the crystal correlation matrix. Although the multiplet occupies all three subsystems, its state consists entirely of bipartite-entangled states and does not present irreducible tripartite entanglement. The existence of zero-configurations in crystal quenches has direct consequences on the dynamics of multipartite entanglement and correlation. In particular, it can further increase the time delay after which tripartite entanglement starts building up in the system, and also explain the negative tripartite mutual information observed in \cite{Caceffo:2023hns}.

A diagnostic is provided by the Markov gap \cite{Hayden:2021gno} (see \cite{Akers:2019gcv,Siva:2021cgo,Berthiere:2023bwn,Balasubramanian:2025jhq} for further studies), whose positivity signals irreducible tripartite entanglement in pure tripartitions \cite{Zou:2020bly}. We build its quasiparticle picture, and obtain explicit expressions for crystal quenches. The Markov gap exhibits nontrivial dynamics, including a delayed growth, nonmonotonic behavior, and a strong dependence on the initial state, which we characterize analytically, confirming recent numerical investigations \cite{Berthiere:2024sio}.
We further show that, for crystal multiplets, zero-configurations are the only nontrivial tripartitions with vanishing Markov gap, hence establishing it as a faithful witness of irreducible tripartite entanglement.

This paper is organized as follows. Section \ref{Sec:setup} introduces our crystal quenches and the correlated multiplets they produce, and briefly review the Markov gap. Section \ref{sec:cms-structure} investigates the multipartite entanglement structure of crystal multiplets and identifies zero-configurations. Their consequences for the onset time and tripartite mutual information are discussed in Sec.\,\ref{Sec:consequences}. Section \ref{Sec:dynamics} develops the quasiparticle picture of the Markov gap and derives its long-time behavior. Both infinite and finite systems are investigated. We conclude in Sec.\,\ref{Sec:Conclusion} with a summary of our results, and give an outlook on future study. Finally, some technical details and further calculations can be found in the six appendices that complete this work.

\section{Crystal quenches, correlated multiplets, and Markov gap}
\label{Sec:setup}

In this section, we introduce our crystal quench protocol, and describe the associated multiplets of correlated quasiparticles. We then review the Markov gap as a probe of irreducible tripartite entanglement and how to compute it for fermionic Gaussian states.

\subsection{Crystal quenches}

We prepare a fermionic chain of length $L$ at $t=0$ in the state $\ket{\psi_0}$. At time $t>0$, we let the initial state evolve as $\ket{\psi(t)}=e^{-itH}\ket{\psi_0}$ under the tight-binding Hamiltonian
\bea\lb{HXX}
H = -\frac12\sum_{i=1}^L\Big(c^\dagger_i c_{i+1} + c^\dagger_{i+1} c_i \Big),
\eea
with $c_i$ standard fermionic operators. Hamiltonian \eqref{HXX} is diagonalized through a Fourier transform
\bea\lb{FT}
d_k=\frac{1}{\sqrt L}\sum_{j=1}^L e^{ikj}c_j\spp, \quad\; k=\frac{2\pi }L \big(\mathbb{Z}\cap [1,L]\big)\spp,
\eea
yielding
\bea\lb{Hdiag}
H=\sum_k\eps(k) d_k^\dagger d_k\spp,\quad\; {\rm with}\quad 
\eps(k)=-\cos k\spp,
\eea
where $d_k^\dagger \,(d_k)$ is interpreted as creating (annihilating) a quasiparticle of momentum $k$ and energy $\eps(k)$. 

In this paper, we mainly focus on a class of initial states with periodic patterns, namely crystal states (see, e.g., \cite{Berthiere:2024sio,Caceffo:2023hns,Bastianello:2018fvl}), which are defined as 
\begin{equation}
|C_{p,q}\rangle = \prod_{i=0}^{L/p-1}\hspace{-2pt}\Bigg (\prod_{l=1}^{q} c_{pi+l}^\dagger\Bigg)|0\rangle = |\underbrace{1\hdots1}_{q}\underbrace{0\hdots0}_{p-q}1 \dots\rangle,
\label{CrystalStates}
\end{equation}
where $0<q<p$ and $L$ is divisible by $p$. The crystal state $|C_{p,q}\ra$ is Gaussian, invariant under translation of $p$ sites, and has filling $\nu=q/p$ with $q$ fermions in each unit cell of $p$ sites. 
For $p=2$ and $q=1$, it reduces to the well-known Néel state. We assume $p>2$ throughout this work.

\subsection{Correlated multiplets}

The eigenstates of the tight-binding Hamiltonian have the form
\bea
\qquad|\Psi_{k_1,\ldots,k_N}\rangle&=d_{k_1}^{\dagger}\cdots d_{k_N}^{\dagger}|0\rangle,
\qquad 0\le N\le L,\quad k_i=\frac{2\pi}{L}n_i,\quad
1\le n_1<\cdots<n_N\le L\,.
    \label{Eigenstates}
\eea
The overlap between \eqref{CrystalStates} and \eqref{Eigenstates} can be computed as
\begin{equation}
    \langle C_{p,q}|\Psi_{k_1,\ldots,k_N}\rangle=\frac{\delta_{N,qL/p}}{L^{N/2}}\det\spm\big\{ e^{-\ii k_j(ip+l)}\big\}_{(i,l),j=1,\dots N} \,, \quad\; 
    (i,l)=\mathbb{Z}^2\cap\Big([0,L/p-1] \times [1,q]\Big) \,,
\end{equation}
such that eigenstates contributing to the dynamics, i.e.~those with non-zero overlap, should satisfy
\bea
    N=\frac{qL}{p}\,, \quad\text{and} \quad \Big|\big\{a,b\in \{1,...,L/p\}:k_a\equiv k_b \text{\;\,(mod }2\pi/p) \big\}\Big|= q \,.
    \label{Conditions}
\eea
This means that for any given $k$, exactly $q$ quasimomenta in $\big \{k,k-\frac{2\pi}{p},\dots ,k-(p-1)\frac{2\pi}{p}\spm \big\}$
must appear in $|\Psi_{k_1,\dots ,k_N}\rangle$. This can be enforced by choosing $k\in(\pi- 2\pi/p,\pi]$.
Such constraints produce $p$-plets of correlated quasiparticles formed by $q$ particles and $p-q$ holes in the eigenstates contributing to the dynamics. The quasiparticles are described \cite{Caceffo:2023hns} by new fermionic operators $\eta_j^\dagger(k)$ as
\begin{equation}
\eta^\dagger_j(k)=d^\dagger_{k-(j-1)2\pi/p}\,, \qquad \varepsilon_j(k)=\varepsilon \Big(k-(j-1)\frac{2\pi}{p}\Big),
\qquad v_j(k)=\partial_k\varepsilon_j(k)=\sin\spm\Big(k-(j-1)\frac{2\pi}{p}\Big),
\label{QPPDef}
\end{equation} 
living in the restricted Brillouin zone $\mathcal{B}=(\pi- 2\pi/p,\pi]$ with energy $\varepsilon_j(k)$ and group velocity $v_j(k)$. Here, $j = 1,\cdots,p$ runs over the quasiparticles species. Using this, the tight-binding Hamiltonian can be put in the form
\bea\lb{HdiagQP}
H=\int_\mathcal{B}\frac{dk}{2\pi}\sum_{j=1}^p \eps_j(k) \eta_j^\dagger(k) \eta_j(k)\spp.
\eea

Since the initial states we consider are Gaussian, time evolution after the quench is completely characterized by the two-point correlation function. In the space of quasiparticle species, the correlation matrix $\mathcal{C}$ calculated on the initial states considered in this paper is block-diagonal \cite{Bastianello:2018fvl,Bertini_2018}
\begin{equation}
\mathcal{C}(k,k') = \langle \psi_0 | \Gamma(k) \Gamma^\dagger(k') |\psi_0 \rangle=\delta_{k,k'}\mathcal{C}(k) \,, \qquad \Gamma^\dagger(k)=\big( \eta_1^\dagger(k),\dots , \eta_p^\dagger(k),\eta_1(k),\dots ,\eta_p(k) \big),
\end{equation}
such that at fixed quasimomentum $k$ we have
\begin{equation}\lb{correlmat}
\mathcal{C}(k)=\begin{pmatrix}
1-G^T(k) & F(k) \\ F^\dagger(k) & G(k)
\end{pmatrix}\spm\spm,
\qquad
 G_{ij}(k)=\langle \psi_0 | \eta_i^\dagger(k) \eta_j(k) |\psi_0\rangle, \quad F_{ij}(k)=\langle \psi_0 | \eta_i(k) \eta_j(k) |\psi_0\rangle.
\end{equation}
Note that for our initial states, the anomalous correlator $F$ vanishes so we only need to consider the matrix $G$. The correlation matrix is the main building block for constructing a quasiparticle picture of entangled multiplets.
At fixed $k$, restrictions of $G$ to selected species determine the entanglement carried by one multiplet, while the velocities determine how that contribution is distributed in space and time. 

In this paper, we are interested in the dynamics of irreducible tripartite entanglement, focusing on the time evolution of the Markov gap as a probe.
We label the quasiparticles of a multiplet $\mathcal Q=\{1,\ldots,p\}$, with $\Q_X\subset\Q$ those in subsystem $X=A,B,C$, and we denote $G_{XY}$ the restriction of the correlation matrix $G$ to $\Q_X\cup\Q_Y$ while $G_{xy}=(G_{ij})_{i\in\mathcal Q_X,j\in\mathcal Q_Y}$ denotes the rectangular cross block of $G_{XY}$. A partition of a multiplet is $\cP=\{\Q_A,\Q_B,\Q_C\}$. 

\subsection{Markov gap and irreducible tripartite entanglement}

The Markov gap is a computable witness of irreducible tripartite entanglement. It can be defined as the difference between the reflected entropy \cite{Dutta:2019gen,Jeong:2019xdr,Kudler-Flam:2020url,Bueno:2020vnx,Li:2020ceg,Berthiere:2020ihq,Bueno:2020fle,Akers:2022max,Vasli:2022kfu,Afrasiar:2022fid,Sohal:2023hst} and the mutual information. 
Consider a pure state $|\Psi\ra$ divided into three subsystems $A$, $B$ and $C$, and its reduced density matrix on $A\cup B$, $\rho_{AB} = \Tr_C|\Psi\ra\spm\la\Psi|$. The canonical purification of $\rho_{AB}$ is given by the square root of the density matrix taken as a state in a doubled Hilbert space $\Hh_A\otimes \Hh_B\otimes\Hh_{A^*}\spm\otimes \Hh_{B^*}$. The reflected entropy $S_R(A:B)$ is the entanglement entropy of $AA^*$ in the canonical purification, $S_R(A:B) = S(AA^*)$ \cite{Dutta:2019gen}. The Markov gap $M(A:B)$ is defined as the difference
\bea\lb{MGdef}
M(A:B)=S_R(A:B) - I(A:B)\ge 0\,,
\eea
where $I(A:B)=S(A)+S(B)-S(A\cup B)$ is the mutual information between $A$ and $B$, and $S(X)=-\Tr\rho_X\spm\log\rho_X$ is the entanglement entropy of subsystem $X$.
As shown in \cite{Zou:2020bly}, the Markov gap vanishes if and only if the tripartite pure state $|\Psi\ra$ is a sum of triangle states (SOTS), i.e.~if each local Hilbert space admits a decomposition $\Hh_X =\bigoplus_i \Hh_{X_L^i}\otimes\Hh_{X_R^i}$ $(X=A,B,C)$, such that, up to local isometries,
\bea\lb{SOTS}
  \ket{\Psi}
  &=\sum_i\sqrt{p_i}\,
    \ket{\psi_i}_{A_R^iB_L^i}
    \ket{\psi_i}_{B_R^iC_L^i}\ket{\psi_i}_{C_R^iA_L^i},
  \eea
where $\ket{\psi_i}_{X_R^iY_L^i}$ is a pure state in $\Hh_{X_R^i}\otimes\Hh_{Y_L^i}$, and $\sum_ip_i=1$.
A triangle state is a trivial SOTS with only one sector, and consists entirely of pairwise distributed bipartite-entangled states. States with positive Markov gap thus present irreducible tripartite entanglement that cannot be reduced to bipartite entanglement with local unitaries \cite{Zou:2020bly}. Note that while in general $M(A: B)\neq M(B : C)\neq M(C : A)$, if one vanishes, then all of them do. 
However, a vanishing Markov gap does not imply that the state has no irreducible tripartite entanglement---for example, the GHZ state is a nontrivial SOTS. 

For  Gaussian fermionic states, entanglement entropies can be calculated from restrictions of the correlation matrix. 
The contribution to $S(A)$ of one multiplet is
\begin{equation}
s_A(k,\Q_A)=-\tr\,\C_A \log \C_A=-\Tr \big ( G_A \log G_A + (1-G_A) \log (1-G_A) \big),
\label{EE}
\end{equation}
where $\C_A$ is the restriction of $\C$ to the subset of quasiparticles present in $A$, and similarly for $G_A$. The second equality holds for number-preserving Gaussian fermionic states. 
We refer to this contribution as an entropy density in the quasiparticle picture, and define
\bea\label{EEdensity}
s(x)=-x\log x-(1-x)\log(1-x)\,.
\eea
For $|C_{p,1}\rangle$, a subsystem containing $|\mathcal Q_A|$ species has a rank-one correlation matrix with nonzero eigenvalue $|\mathcal Q_A|/p$, hence $s_A=s(|\mathcal Q_A|/p)$. 
Note that for $p>2$ and $1<q<p-1$, $s_A(k,\Q_A)$ is in general not directly related to the thermodynamic entropy of the generalized Gibbs ensemble (GGE) that describes the steady state \cite{Calabrese:2016psf}, as it is the case when only entangled pairs of quasiparticles are produced after the quench \cite{Alba:2017ekd,Alba:2018hie}. In particular, the GGE only captures diagonal terms of $G(k)$. Although this is sufficient for obtaining the entanglement entropy time evolution from $|C_{p,1}\ra$, for generic $p>2$, $1<q<p-1$ the full correlation matrix needs to be evaluated, not only its diagonal elements. This breaks the relationship between the GGE thermodynamic entropy and the multiparty entanglement content (see \cite{Caceffo:2023hns} for more details).

For number-conserving Gaussian fermionic states, the reflected entropy is computed from the reflected correlation matrix $\mathbf{G}_{AA^*}$ defined as \cite{Bueno:2020vnx} (see also \cite{Berthiere:2023gkx}) 
\begin{equation}
\mathbf{G}_{AA^*}=\begin{pmatrix}
G_{A} & \sqrt{G_{AB}(1-G_{AB})}\big|_A \\
\sqrt{G_{AB}(1-G_{AB})}\big|_A & 1-G_{A}
\end{pmatrix} \spm,
\label{Gbold}
\end{equation}
where $G_{AB}$ is the restriction of the original correlation matrix $G$ to the subset of quasiparticles $\Q_A\cup\Q_B$.
The contribution of a multiplet to $S_R(A:B)$ is given by
\begin{equation}
s^R_{A:B}(k,\cP)\equiv s_{AA^*}(k,\cP) =-\Tr\spp\big[\mathbf{G}_{AA^*} \log\mathbf{G}_{AA^*}+(1-\mathbf{G}_{AA^*})\log(1-\mathbf{G}_{AA^*})\big].
\label{RefEntDens}
\end{equation}
The Markov gap density follows as
\begin{equation}
\lb{MGdensdef}
\begin{aligned}
m_{A:B}(k,\cP)
&=s_{AA^*}(k,\cP)-s_{A}(k,\mathcal{Q}_A)-s_{B}(k,\mathcal{Q}_B)+s_{AB}(k,\mathcal{Q}_{AB})\,.
\end{aligned}
\end{equation}
A multiplet can contribute to the Markov gap if at least one quasiparticle is present in each subsystem $A, B, C$, i.e.~for nontrivial configurations with $|\Q_X|\ge 1$ for all $X=A,B,C$.
Understanding the entanglement structure of a system after a quantum quench in the quasiparticle picture amounts to understanding the entanglement structure of the multiplets. This is what we do in the next section.

\section{Entanglement structure of crystal multiplets}
\label{sec:cms-structure}

The entanglement carried by a crystal multiplet depends on how its species are distributed among the subsystems. In particular, a multiplet shared by three subsystems may not carry irreducible tripartite entanglement. We identify \textit{zero-configurations} for which correlations can be resolved into independent bipartite resources. For small multiplets, we obtain a complete picture of their tripartite entanglement structure. More generally, we use the Markov gap as a probe of irreducible tripartite entanglement. Finally, we discuss the extension to more than three parties.

\subsection{Multiplet state}
\label{subsec:cms-conventions}

For a general crystal state \eqref{CrystalStates}, the correlation matrix $G(k)$ in the quasiparticle basis is independent of $k$,
\bea\label{Gmat}
G_{ij}=\frac1p\sum_{l=1}^{q}e^{2\pi\ii l(i-j)/p},
\qquad G_{ii}=\frac qp.
\eea
It is the row Gram matrix $G=VV^\dagger$ of the $p\times q$ rectangular Vandermonde matrix $V_{jl}=p^{-1/2}e^{2\pi\ii lj/p}$, $l=1,\ldots,q$.
At fixed $k$, a crystal multiplet is the normalized Slater state
\begin{equation}
 |\Psi_{p,q}\rangle =\prod_{l=1}^{q}\hspace{-2pt}
 \left[ \sum_{X=A,B,C}\sum_{j\in\mathcal Q_X} \hspace{-2pt}V_{jl}^*\eta_j^\dagger(k)\spm\right]\hspace{-3pt}|0\rangle ,
 \label{eq:cms-original}
\end{equation}
where $\mathcal Q=\mathcal Q_A\cup\mathcal Q_B\cup\mathcal Q_C =\{1,\ldots,p\}$, and $|0\rangle$ is the vacuum of all $p$ modes so that $\eta_j(k)|0\rangle=0$. The $p$ species label the available modes, while the state has total fermion occupation $q$. 
We may choose a representation of the kets in terms of subsystem order $A,B,C$, with increasing species
labels within each party, $\mathcal Q_X=\{j_{X,1}<\cdots<j_{X,|\mathcal Q_X|}\}$, 
\bea
 |\boldsymbol n_A;\boldsymbol n_B;\boldsymbol n_C\rangle
 =\prod_{X=A,B,C}\left[\prod_{\alpha=1}^{|\mathcal Q_X|} (\eta_{j_{X,\alpha}}^\dagger)^{n_{X,\alpha}} \right]\hspace{-3pt}|0\ra .
 \label{eq:cms-order}
\eea
Then up to local unitaries, \eqref{eq:cms-original} can be written as
\begin{equation}
 |\Psi_{p,q}\rangle\simeq
 \prod_{l=1}^{q}\left[\sum_{X=A,B,C} \sum_{\alpha=1}^{r_X}
  \sqrt{\lambda_{X,\a}}(v_{X,\alpha})_l\eta_{j_{X,\alpha}}^\dagger \right]\hspace{-3pt}|\mathbf{0};\mathbf{0};\mathbf{0}\rangle,
 \label{eq:cms-normal}
\end{equation}
where $(V_X^\dagger V_X)v_{X,\alpha}=\lambda_{X,\a}v_{X,\alpha}\;(\lambda_{X,\alpha}>0)$ and $r_X\equiv\operatorname{rank}V_X=\min(|\mathcal Q_X|,q)$, with $V_X$ being the restriction of $V$ to the rows $j\in\mathcal Q_X$.

\subsection{Zero-configurations}
\label{subsec:cms-zero}

Let $g=\gcd(p,q)$ and $d=p/g$. Then for $i\neq j$ and $g>1$, off-diagonal zeros appear in the correlation matrix at specific positions given by
\bea
G_{ij}=0 \;\iff\; i\equiv j \;\;(\text{mod}\; d)\,.
\label{ZeroCondition}
\eea
Equivalently, $G_{ij}= 0$ if $i$ and $j$ are distinct elements of the same residue class $R_r$,
\bea
R_r = \big\{r,r+d,\dots,r+(g-1)d\big\}, \quad \text{for some}\quad  r=1,\dots,d\,.
\eea
The quasiparticles thus split into $d$ residue classes of size $g>1$, i.e.~$\cup_{r=1}^d R_r=\Q$. In particular, each $R_r$ contains $g$ mutually uncorrelated quasiparticle species, whereas any two quasiparticles belonging to different residue classes have nonzero correlations.  
We call a partition $\cP$ of a multiplet into three nonempty sets a
\emph{zero-configuration} if 
\bea
\exists X\ne Z,\ r:\ \mathcal Q_X\cup\mathcal Q_Z\subseteq R_r
\quad\Longleftrightarrow\quad
\exists X\ne Z:\ G_{xz}=0\,,
\eea
and denote by $\cP_0$ the set of all zero-configurations for a given $(p,q)$. Note that for $g=1$, every residue class has one element and zero-configurations are absent.

For a zero-configuration $\cP\in\cP_0$, quasiparticles in $X\cup Z$ are not correlated and implies the product marginal $\rho_{XZ}=\rho_X\otimes\rho_Z$. Purity of the full multiplet state $|\Psi_{p,q}^\cP\rangle$ then yields a Hilbert-space decomposition of the remaining subsystem $\Hh_Y=\Hh_{Y_L}\otimes\Hh_{Y_R}$ such that
\begin{equation}\label{DegenerateTriangle}
|\Psi_{p,q}^\cP\rangle=|\psi\rangle_{XY_L}\otimes|\psi\rangle_{Y_RZ}\,,
\end{equation}
up to local isometries on $Y$.
This state consists entirely of pairwise distributed bipartite-entangled states, and thus presents no irreducible tripartite entanglement. Consequently, the Markov gap is zero
\begin{equation}\label{ZeroSufficiency}
\cP\in\cP_0\ \Longrightarrow\ m_{X:Z}(\cP)=0\,.
\end{equation}

For arbitrary pure number-conserving fermionic Gaussian states, we show in Appendix \ref{vanishingMG} that
\bea
M(X:Z)=0 \;\;\iff\;\; G_{ab}G_{bc}=0\,, \quad G_{ca}G_{ab}=0\,, \quad  G_{bc}G_{ca}=0\,.
\eea
For crystal multiplets, we further show in Appendix~\ref{vanishingMG} that a vanishing Markov gap density necessarily implies that at least one cross-correlation block vanishes, which yields the other direction of the implication in \eqref{ZeroSufficiency}. Hence the Markov gap density vanishes if and only if a multiplet is in a zero-configuration,
\bea
\cP\in\cP_0 \iff m_{X:Z}(\cP)= 0 \iff  \text{no irreducible tripartite entanglement.} 
\eea
Empty subsystems give a vanishing gap trivially and are excluded from the classification.
We thus conclude that for crystal multiplets, only zero-configurations have trivial tripartite entanglement, and that a zero Markov gap certifies the lack of irreducible tripartite entanglement.

\subsection{The $q=1$ multiplets: W-type entanglement}
\label{subsec:cms-q1}

For initial crystal states with $q=1$, there are $p$ modes but exactly one occupied fermion per multiplet such that
\bea
|\Psi_{p,1}\ra \simeq \sqrt{\nu_A}|100\ra + \sqrt{\nu_B}|010\ra +  \sqrt{\nu_C}|001\ra\spp,
\eea
where $\nu_X=|\Q_X|/p$, and $\simeq$ means equality up to local unitaries. This state is a weighted W state, with weights determined entirely by the number of quasiparticles in each subsystems. Hence only W-type entanglement can be produced after a $q=1$ quench in the hydrodynamic regime. The Markov gap is easily obtained as
\bea
 m_{A:B}^{(p,1)}= 2s\!\left[ \frac12\left(1+ \sqrt{1-\frac{4\nu_A \nu_B}{\nu_A+\nu_B}}\right)\right]-s(\nu_A)-s(\nu_B) +s(\nu_C)\spp,
 \label{eq:cms-q1-gap}
\eea
where $s(x)$ is defined in \eqref{EEdensity}.

\subsection{The $C_{4,2}$ multiplet: a complete tripartite classification}
\label{subsec:cms-c42}

For $C_{4,2}$, there are 36 nontrivial tripartitions, among which 12 are zero-configurations and 24 are irreducible tripartite entangled. Every nontrivial tripartitions has full-rank $(2,4,2)$, where the ranks refer to the reduced density matrices on $X,Y,Z$ ($X,Y,Z\in\{A,B,C\}$). 

Multiplets in a zero-configuration contain two maximally entangled pairs, $X$--$Y$ and $Y$--$Z$, with $|\Q_Y|=2$. The singleton parties have entropy $\log 2$ while the doublet party has entropy $2\log 2$, and $m_{A:B}=m_{A:C}=m_{B:C}=0$. These states are entangled across every bipartition while being assembled entirely from bipartite resources.
For example, the multiplet's state corresponding to the configuration $\cP=\{\{1\},\{2,4\},\{3\}\}$ with a doublet in $B$,
 \bea
|\Psi^{\cP}_{4,2}\rangle\simeq
\frac12\big(|1;00;1\ra+|1;01;0\ra +|0;10;1\ra +|0;11;0\ra\big)=\left(\frac{|01\ra+|10\ra}{\sqrt2}\right)_{AB_L}\otimes\left(\frac{|01\ra+|10\ra}{\sqrt2}\right)_{B_RC},
\eea
contains two Bell pairs, one shared between $A$ and the effective mode $B_L$, and the other between the effective mode $B_R$ and $C$. The labels $B_L$ and $B_R$ refer to their respective Fock spaces, rather than to the original species $2$ and $4$. The rotation $2b_{L/R}^\dagger=(-1\mp i)\eta_2^\dagger+(-1\pm i)\eta_4^\dagger$ defining these modes acts entirely within $B$.

For non-zero-configurations, the reason all these states have irreducible tripartite entanglement is as follows. Let $Y$ denote the party containing two species, and let $X$ and $Z$ denote the two singleton parties. Their reduced density matrices are full rank: $\rank\spp\rho_X=2$, $\rank\spp\rho_Y=4$, $\rank\spp\rho_Z=2$.
A decomposition into independent bipartite states would require one $X$--$Y$ pair and one $Y$--$Z$ pair, which would imply $I(X:Z)=0$. Instead we have $I(X:Z)>0$, hence such a decomposition is impossible.
For example, the multiplet's state corresponding to the configuration $\mathcal{T}=\{\{1\},\{3,4\},\{2\}\}$ with a doublet in $B$ is, up to local rotations,
\bea
|\Psi^\mathcal{T}_{4,2}\rangle \simeq
\frac12\big(|1;01;0\ra+|0;10;1\ra\big)
+\frac{1}{2\sqrt2}\big(|0;11;0\ra+|1;00;1\ra-|1;10;0\ra-|0;01;1\ra\big),
\eea
and carries irreducible tripartite entanglement, $m_{A:B}(\mathcal{T})>0$. All non-zero-configuration states take a similar form.

\subsection{Larger multiplets can contain several tripartite resources}
\label{subsec:cms-larger}

For a general $(p,q)$ multiplet, there are $3(3^{p-1}-2^p+1)$ nontrivial (labeled) tripartitions, including $3d(3^{g}-2^{g+1}+1)$ zero-configurations. 
For large multiplets, non-zero-configuration need not be one indivisible entangled state on all its local modes. For example, a partition of a $C_{6,2}$ multiplet can be a product of two tripartite W states. More generally, the structure of entanglement becomes richer as multiplets get larger.

\subsection{Extension to more than three parties}
\label{subsec:cms-multipartite}

Our discussion can be extended to $N$ parties for $p\ge N$, $\Q=\bigcup_{a=1}^{N}\Q_{X_a}$. For crystal multiplets with $q=1$, nontrivial configurations are generalized W states
\bea
 |\Psi_{p,1}^{(N)}\rangle\simeq
 \bigg[\sum_{a=1}^{N} \sqrt{\nu_{X_a}}
 \eta_{j_{X_a,1}}^\dagger \bigg]|0\ra.
 \label{eq:cms-N-W}
\eea
Each party has local rank two, and every nontrivial bipartition has positive entropy. 
\pagebreak

One can introduce generalized zero-configurations satisfying 
$\Q_{X_{i_1}}\cup\cdots\cup\Q_{X_{i_K}}\subseteq R_r$ with $K\leq N-1$ for which
$\rho_{{X_{i_1}}\cup\cdots\cup{X_{i_K}}}=\bigotimes_{a=1}^{K}\rho_{X_{i_a}}$.
As an example, take $C_{4,2}$ and the partition $A=\{1\}$, $B=\{2\}$, $C=\{3\}$, and $D=\{4\}$. Both $\rho_{AC}=\rho_A\otimes\rho_C$ and $\rho_{BD}=\rho_B\otimes\rho_D$ are product marginals. Nevertheless, the pure multiplet's state is entangled across every bipartition. Since each subsystem has rank two, independent lower-partite entanglement would contradict this. Indeed, two independent resources entangling the same party would give its reduced state rank at least $2\times2=4$, while resources supported on disjoint groups of parties would make the state separable across a bipartition. This state therefore presents irreducible quadripartite entanglement, as do every nontrivial configurations of $C_{4,2}$. For crystal multiplets with $p=6$, the story is different. There exist four-party configurations that reduce to independent tripartite states (for $q=2$), and others that reduce entirely to bipartite entangled states (for $q=3$). We leave the detailed study of multipartite entanglement of crystal quenches to the interested reader.

\section{Consequences of zero-configurations}
\lb{Sec:consequences}

In the quasiparticle picture, a multiplet is produced at each point in space at $t=0$, with arbitrary quasimomentum $k$. Quasiparticles then spread ballistically through the system with group velocity $v_j(k)=\partial_k\varepsilon_j(k)$ and populate the different subsystems. Each multiplet contributes to the growth of multipartite entanglement or correlation in the system as its quasiparticles are being distributed among the subsystems. 
The existence of zero-configurations in crystal quenches has important repercussions on the dynamics of multipartite entanglement and correlation. In particular, it can increase the time delay $t_e$ after which tripartite entanglement starts building up in the system, and also explain why the initial state $|C_{4,2}\ra$ produces negative tripartite mutual information.

\subsection{Delay of the onset time}

The onset time is (naively) obtained by the fastest entangling pair of quasiparticles as
\begin{equation}
t_e=\frac\ell{\max\spm\big|v_{i}(k)-v_j(k)\big|},
\label{taue}
\end{equation}
where $\ell=\min(\ell_A,\ell_B)$, $v_j(k)$ is given in \eqref{QPPDef} and $k\in (\pi-2\pi/p,\pi]$.  Using cyclic relabeling to unfold the restricted Brillouin zone gives
\bea
\max\spm\big|v_{i}(k)-v_j(k)\big|=2\cos\spm\spm\left(\spm\frac{\pi}{2p} \frac{1-(-1)^p}2\spm\right) \iff |i-j| = 
\begin{dcases}
\spp\frac p2 \,,  & p\,\; \text{even}, \\
\spp\frac{p\pm1}2 \,,  &p\,\; \text{odd},
\end{dcases}
\eea
which reduces to 2 for $p$ even, and to $2\cos(\pi/2p)$ for $p$ odd. For even $p$, the antipodal separation $p/2$ is a zero-configuration if $g$ is even, associated with a zero Markov gap density. In that case, the next fastest contributing pair satisfies $|i-j| = p/2\pm1$. 
The onset time then follows as
\bea
\qquad\qquad\frac{t_e}{\ell}=
\begin{cases}
[2\cos(\pi/2p)]^{-1},&p\ \text{odd},\\
1/2,&p\ \text{even},\ g\ \text{odd},\\
[2\cos(\pi/p)]^{-1},&p\ \text{even},\ g\ \text{even} \;\;\longrightarrow\;\;\text{consequence of zero-configurations}\spp.
\end{cases}
\eea
Zero-configurations can thus further delay the production of irreducible tripartite entanglement in crystal quenches. For example, for the initial state $|C_{4,1}\ra$ we obtain an onset time $t_e^{(4,1)}=\ell/2$, whereas for $|C_{4,2}\ra$ we find $t_e^{(4,2)}=\ell/\sqrt{2}$.
The same delay of the onset time can also be observed for bipartite quantities such as mutual information and fermionic logarithmic negativity of disjoint subsystems (mixed bipartite states).

\subsection{Negative tripartite mutual information}

A negative tripartite mutual information (TMI) signals that mutual information is monogamous, which reflects that correlation shared between three intervals is more than the sum of the pairwise correlations between them. The quasiparticle picture for the TMI after free-fermionic quantum quenches is developed in \cite{Caceffo:2023hns}, where it was shown that a quench from the initial state $|C_{4,2}\ra$ yields a nonpositive TMI in the hydrodynamic regime. This nonpositiveness of TMI can be attributed to the existence of zero-configurations.

The connection follows from writing the TMI in terms of the conditional mutual information $I(X:Y\mid Z)$,
\bea
I_3(A:B:C)=I(A:B)-I(A:B\mid C)=I(B:C)-I(B:C\mid A)\spp.
\eea
Any assignment of the four species to $A,B,C,D$ leaves, among the three species in $A\cup B\cup C$, a complete residue class \{1,3\} or \{2,4\}. Hence two of $A,B,C$ are uncorrelated such that
\bea
i_3(A:B:C)=\displaystyle -i(X:Y\mid Z)\le 0\spp,
\eea
by strong subadditivity.
Since all kinematic weights are nonnegative, the full TMI is thus nonpositive at all times in the hydrodynamic regime. 
Zero-configurations therefore explain the sign of TMI for $|C_{4,2}\ra$ as they control every contributing configuration. For larger multiplets, the existence of zero-configurations is neither sufficient nor necessary in general for the TMI to be nonpositive. For example, $|C_{5,2}\ra$ has no zero-configurations but can still have negative hydrodynamic TMI.

\section{Tripartite entanglement dynamics after crystal quenches}
\lb{Sec:dynamics}

We derive here the quasiparticle picture for tripartite entanglement dynamics in the presence of multiparticle excitations, focusing on three representative examples.
To build the quasiparticle picture of the post-quench time evolution of the Markov gap, one needs two ingredients: $i)$ the contribution of a given partition of a multiplet, and $ii)$ the number of multiplets in that configuration at time $t$. 
The first ingredient is the density of Markov gap \eqref{MGdensdef}, given for a fixed $k$ and a given quasiparticle configuration $\cP$. The second ingredient is encoded in a kinematic weight $\D$ which reads
\begin{equation}\label{GeneralKernel}
\mathcal D(k,\cP,t)=\int_{\mathbb R}dx\spm
\prod_{X=A,B,C}\prod_{j\in\mathcal Q_X}\mathbf1_X\!\left(x+v_j(k)t\right),
\end{equation}
where $\mathbf1_X$ denote the indicator of the corresponding spatial region $X$.
Finally, one must consider all possible quasiparticle partitions, for all quasimomenta $k$ in the restricted Brillouin zone.
The quasiparticle prediction is
\bea
M_{A:B}(t) = \int_\mathcal{B} \frac{dk}{2\pi} \sum_{\cP}m_{A:B}(k,\cP)\D(k,\cP,t) ,
\eea
where the sum is over labeled partitions with all three parties nonempty. This is the leading hydrodynamic expression. For crystal states the densities are independent of $k$.

Unless stated otherwise, we consider a system divided into three parts $A$, $B$ and $C$, with respective lengths $\ell_A$, $\ell_B$ and $\ell_C$. For simplicity, we take $A$ and $B$ adjacent such that $A=[-\ell_A,0]$ and $B=[0,\ell_B]$, and we work in the thermodynamic limit of infinite system size, which we access by letting the length of subsystem $C$ be infinite. We choose the characteristic length scale to be $\ell=\min(\ell_A,\ell_B)$ and we also define the aspect ratio $r=\max(\ell_A/\ell_B, \ell_B/\ell_A)$. In Appendix \ref{App:WStates}, we consider another class of initial states for which a single fermion is in a W--state in each cell.

\subsection{Initial state producing triplets}
\label{Sec:CrystalStates}

\begin{figure}[h!]
\centering
\hspace{-1cm}
\includegraphics[scale=1.4]{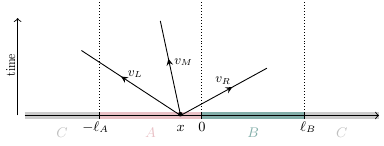}
\caption{Initial setup and quasiparticle propagation. Two regions $A$ and $B$ of respective lengths $\ell_A$ and $\ell_B$ are defined on an infinite region, along with their complementary region $C$. The quasiparticles are created at every point of the lattice at $t=0$, and propagate ballistically through time. They are labeled by their velocities $v_R,v_M$, and $v_L$. Entanglement arises when each quasiparticle of a given triplet is in a different region than the other two.}
\label{Fig:InitSetup}
\end{figure}

We begin by studying the post-quench dynamics for the initial state $|C_{3,1} \rangle=|\cdots100100\cdots\rangle$. According to \eqref{QPPDef}, triplets of quasiparticles are generated throughout the system after the quench, with group velocities
\begin{equation}
v_j(k)=\sin \Big (k -(j-1) \frac{2\pi}{3} \Big), \quad j=1,2,3 \,,
\end{equation}
such that $\sum_j v_j(k)=0$. 
Typical triplets generated after the quench are depicted in Fig.\,\ref{Fig:InitSetup} for a fixed $k$. One may choose to label the quasiparticles by their velocity ordering $v_L(k)\le v_M(k) \le v_R(k)$. A triplet $\{L,M,R\}$ contributes to tripartite entanglement if and only if one quasiparticle is present in each subregion $A$, $B$, $C$. Regions $A$ and $B$ are adjacent finite intervals, while $C$ is their infinite complement. There are thus two possible arrangements that can contribute to tripartite entanglement: $L$ goes into the left component of $C$ or $R$ goes into the right component of $C$, while the two remaining quasiparticles are respectively in $A$ and $B$. To each configuration corresponds a kinematic term, denoted $\mathcal{D}_L$ and $\mathcal{D}_R$, respectively.

\textit{Evaluation of $\spp\mathcal{D}_L$ and $\spp\mathcal{D}_R$} --- We first consider a triplet $\{L,M,R\}$ generated at position $x$ at $t=0$ and whose quasiparticle with velocity $v_L$ goes into the left component of $C$ while $M\in A$ and $R\in B$ (see Fig.\,\ref{Fig:InitSetup}). For tripartite entanglement to arise in this situation, the following constraints must be satisfied:
\bea
-\ell_A-v_M(k)t\le &\,\, x \le -v_M(k) t  \hspace{1.25cm}  \Leftrightarrow\;\Q_A=\{M\},\nn\\
-v_R(k)t\le &\,\, x \le \ell_B-v_R(k) t    \hspace{0.786cm}  \Leftrightarrow\; \Q_B=\{R\},\\\nn
 &\,\, x \le -\ell_A -v_L(k) t                   \hspace{0.545cm}  \Leftrightarrow\; \Q_C=\{L\}.
\eea
For any given $k$, these constraints define a region inside of which all the entangled triplets contribute to the total entanglement. Integrating over $x$ yields
\begin{equation}
\begin{aligned}
\mathcal{D}_L(k,t)&=\max\spm\Big[0,\min\spm\big(-v_M(k)t,\ell_B-v_R(k)t,-\ell_A-v_L(k)t\big) - \max\spm\big(-\ell_A-v_M(k)t,-v_R(k)t\big)\Big].
\end{aligned}
\label{D1MM}
\end{equation}

Similarly, a triplet emitted at position $x$ at $t=0$ whose quasiparticle with velocity $v_R$ goes into the right component of $C$ while $L\in A$ and $M\in B$ contributes to entanglement if the following constraints are satisfied:
\bea
-\ell_A-v_L(k)t\le &\,\, x \le -v_L(k) t  \hspace{1.46cm}  \Leftrightarrow\; \Q_A=\{L\},\nn\\
-v_M(k)t\le &\,\, x \le \ell_B-v_M(k) t    \hspace{0.815cm}  \Leftrightarrow\; \Q_B=\{M\},\\\nn
 \ell_B-v_R(k)t \le&\,\, x                \hspace{3.18cm}  \Leftrightarrow\; \Q_C=\{R\}.
\eea
Upon integrating over all possible positions $x$, one obtains
\begin{equation}
\begin{aligned}
\mathcal{D}_R(k,t) &= \max\spm\Big [0, \min\spm\big (-v_L(k)t ,\ell_B-v_M(k)t\big) - \max\spm\big(-\ell_A-v_L(k)t,-v_M(k)t,\ell_B-v_R(k)t \big)\Big] .
\end{aligned}
\label{D2MM}
\end{equation}
Note that the kinematic weights $\D_{L/R}$ are general for tripartite correlation dynamics following a quench from an initial state producing triplets of quasiparticles. A more in-depth study of these weights is performed in Appendix\,\ref{App:Pheno}.

\medskip

\textit{Density of Markov gap} --- Since $g=1$, no zero-configurations occur. For triplets of quasiparticles, every nontrivial configuration with one species in each region contributes the same nonzero density
\begin{equation}\lb{mgdens31}
m^{(3,1)}_{A:B}=2s\bigg(\frac{3+\sqrt{3}}{6} \bigg) -s(1/3) \,,
\end{equation}
where $s(x)$ is defined in \eqref{EEdensity}.

\medskip

\begin{figure}[t]
\centering
\includegraphics[scale=1]{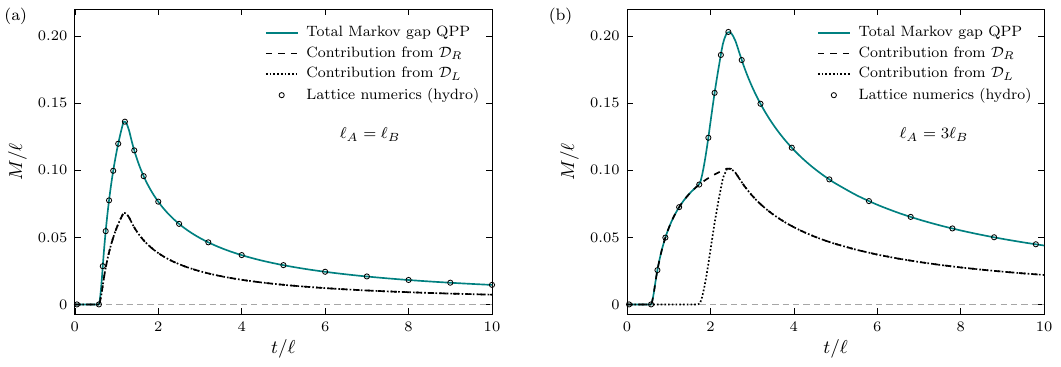}
\caption{Dynamics of the Markov gap following quenches from the initial state $|C_{3,1}\rangle$. Discontinuous lines correspond to individual contributions from $\mathcal{D}_R$ and $\mathcal{D}_L$. Individual points show the exact numerical evaluation in the hydrodynamic limit. (a) Equal susbsystem sizes. (b) Unequal subsystem size $\ell_A =3 \ell_B$.}
\label{Fig:MGapExactEvol}
\end{figure}

\textit{Time evolution of Markov gap} ---
Putting \eqref{D1MM}, \eqref{D2MM} and \eqref{mgdens31} together, we obtain 
\begin{equation}
M_{A:B}(t)=m^{(3,1)}_{A:B}\int_{\pi/3}^\pi \frac{dk}{2\pi} \big( \mathcal{D}_L(k,t)+\mathcal{D}_R(k,t) \big) \spp.
\label{MarkovGap}
\end{equation}
In the hydrodynamic limit of long times and large intervals (with their ratio fixed), the quasiparticle prediction \eqref{MarkovGap} is in perfect agreement with exact lattice results \cite{Berthiere:2024sio}, see Fig.\,\ref{Fig:MGapExactEvol}. As the density of Markov gap is independent of $k$, all nontrivial time-dependence comes from propagation. 
Note that this result is straightforwardly extended to the Rényi-generalized Markov gap  \cite{Berthiere:2023bwn}.

Interestingly, the integral in \eqref{MarkovGap} can be performed analytically, which allows for a detailed analysis of the time evolution of the Markov gap---see Appendix \,\ref{App:ExactDerivation}. First, as was observed in \cite{Berthiere:2024sio}, the Markov gap starts to evolve after a time delay $t_e=\ell/\sqrt3$. 
At time $t\sim t_e$, the Markov gap rapidly increases as $(t-t_e)^{3/2}$, signaling a production of irreducible tripartite entanglement. If $\ell_A\neq\ell_B$, a plateau develops as one of the subsystems becomes larger than the other, followed by a second growth of entanglement at time $t_b=r\ell/\sqrt3$ (note that $t_b=t_e$ for $\ell_A=\ell_B$). For $\ell_A=\ell_B$, the Markov gap reaches its maximum value at $t_{\rm max}=8\ell/(3\sqrt{5})$. The Markov gap then decreases and vanishes as $(t/\ell)^{-1}$ in the long-time limit, thus displaying an entanglement barrier. This decay behavior is universal for all crystal quenches, as we show in Subsec.\,\ref{longtime}.
Entanglement typically increases when more momentum sectors display a growth of entanglement than the ones where entanglement decays, as is shown in Appendix\,\ref{App:ExactDerivation}. In particular, we can identify a time $t_D=(2\ell/3)\sqrt{1+r+r^2}<t_{\rm max}$ at which some of the momentum sectors begin to decay before the peak of entanglement. This signals an impending population inversion between the increasing and decreasing momentum sectors, which occurs at $t=t_{\rm max}$. However, since every sector does not increase or decrease equally, we observe an inertia that explains why $t_D<t_{\rm max}$. Finally, closed-form evaluation of \eqref{MarkovGap} revealed $r$-dependent behavior beyond the one observed Fig.\,\ref{Fig:MGapExactEvol}.

\subsection{Initial states producing quadruplets}

Let us now consider initial crystal states producing multiplets of four quasiparticles, $|C_{4,1} \rangle$ and $|C_{4,2} \rangle$, with velocities
\begin{equation}\lb{vel41}
v_j(k)=\sin \Big (k-(j-1)\frac{\pi}{2} \Big) \,, \quad j=1,\dots,4 \,.
\end{equation}
With multiplets of four quasiparticles, phenomenology gets richer as a quadruplet can be distributed in several inequivalent ways among the three subsystems, each configuration potentially contributing differently to tripartite entanglement. Since the quasiparticles velocities are identical for $|C_{4,1} \rangle$ and $|C_{4,2} \rangle$, one might expect their dynamics to be the same. However, comparison between the two quenches exposes the effect of internal multiplet structure: for $|C_{4,1} \rangle$ we have $g=\gcd(4,1)=1$ and no zero-configurations can occur, while for $|C_{4,2} \rangle$ we have $g=\gcd(4,2)=2$ such that zero-configurations are present.

\begin{figure}[b]
\centering
\includegraphics[scale=1.]{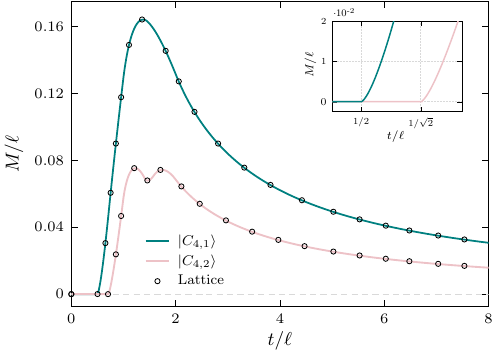}
\caption{Dynamics of the Markov gap following a quench of initial states $|C_{4,1}\rangle$ and $|C_{4,2}\rangle$ in the tight-binding chain with equal subsystem sizes $\ell_A=\ell_B$. Full lines represent the quasiparticle picture while individual dots show the numerical evaluations. The inset presents a close-up around time $t=1/2$ and $t=1/\sqrt{2}$, showing the onset time for the initial states. }
\label{Fig:MG4}
\end{figure}

Since the velocities \eqref{vel41} are distributed in pairs of opposite signs, only two of them are independent. The Brillouin zone can be further reduced on $k\in(\pi/2,3\pi/4)$, where $v_3=-v_1<v_4=-v_2<v_2<v_1$. Given these velocities ordering, there are seven possible quasiparticle configurations with all three subsystems nonempty: $BACC$, $CCAB$, $BBCA$, $CBAA$, $CBCA$, $BACA$, $CBAB$ (species are taken in the order 1,2,3,4). The last three are zero-configurations for $q=2$. The rest of the Brillouin zone is related by exchanging species $1\leftrightarrow2$ and $3\leftrightarrow4$. Following the same method as for $|C_{3,1}\rangle$, we obtain 
\begin{equation}\lb{MG41}
\begin{aligned}
M_{A:B}(t)&=2 \int_{\pi/2}^{3\pi/4} \frac{dk}{2\pi} \Big(m^{(4,q)}_2\mathcal{D}_2^{(q)}(k,t) + m^{(4,q)}_3 \mathcal{D}_3^{(q)}(k,t)\Big) \spp,
\end{aligned}
\end{equation}
where $m_{|\Q_{AB}|}^{(4,q)}$ denotes the nonzero density for $|\Q_{AB}|=2,3$. For $|C_{4,1}\rangle$ we have
\bea
\mathcal{D}_2^{(1)}(k,t)=
 &\spp\max\spm\big[0,\min(\ell_B-v_1 t,-v_2 t,-\ell_A+v_2 t)-\max(-v_1 t,-\ell_A-v_2 t)\big]\nn\\
     &+\max\spm\big[0,\min(\ell_B+v_2 t,v_1 t\big)-\max(\ell_B-v_2 t,v_2 t,-\ell_A+v_1 t)\big]\\\nn
   &+\max\spm\big[0,\min(\ell_B-v_2 t,v_2 t,-\ell_A+v_1 t\big)-\max(\ell_B-v_1 t,-v_2 t,-\ell_A+v_2 t)\big], \\
\mathcal{D}_3^{(1)}(k,t) =
&\spp\max\spm\big[0,\min(\ell_B-v_1 t,v_2 t,-\ell_A+v_1 t) -\max(-v_1 t,-\ell_A+v_2 t)\big] \nn\\
  &+\max\spm\big[0,\min(\ell_B-v_2 t,v_1 t)-\max(\ell_B-v_1 t,v_2 t,-\ell_A+v_1 t)\big]\\\nn
  &+\max\spm\big[0,\min(\ell_B-v_2 t,v_2 t)-\max(\ell_B-v_1 t,-v_2 t,-\ell_A+v_1 t)\big].
\eea
The first term of $\mathcal D_3^{(1)}$ combines configurations $BACA$ and $BBCA$. These configurations have equal densities for $|C_{4,1}\ra$, but only the latter contributes for $|C_{4,2}\ra$. Its weights must therefore be regrouped:
\begin{align}
\mathcal D_2^{(2)}(k,t)=&
\spp\max\spm\big[0,\min(\ell_B-v_1 t,-v_2 t,-\ell_A+v_2 t)-\max(-v_1 t,-\ell_A-v_2 t)\big]\nonumber\\
&+\spp\max\spm\big[0,\min(\ell_B+v_2t,v_1 t)-\max(\ell_B-v_2 t,v_2 t,-\ell_A+v_1 t)\big],\label{QuadD22}\\
\mathcal D_3^{(2)}(k,t)=&
\spp\max\spm\big[0,\min(\ell_B-v_1 t,v_2 t,-\ell_A+v_1 t)-\max(-v_2 t,-\ell_A+v_2 t)\big]\nonumber\\
&+\spp\max\spm\big[0,\min(\ell_B-v_2 t,v_2 t)-\max(\ell_B-v_1 t,-v_2 t,-\ell_A+v_1 t)\big] .\label{QuadD32}
\end{align}
These expressions retain $BACC,CCAB$ and $BBCA,CBAA$, respectively. The two nonzero densities for $C_{4,1}$ are
\begin{align}
m_2^{(4,1)}&=2 s\bigg(\frac{2 + \sqrt{2}}4\bigg) + s(1/2) - 2 s(1/4),\\
m_3^{(4,1)}&=2 s\bigg(\frac{3 + \sqrt{3}}6\bigg) -s(1/2).
\end{align}
For $|C_{4,2}\ra$ they are
\begin{align}
m_2^{(4,2)}&=4s\bigg(\frac{2+\sqrt2}{4}\bigg)-2s(1/2),\\
m_3^{(4,2)}&=2s(1/4)-2s\bigg(\frac{2+\sqrt2}{4}\bigg).
\end{align}
Our quasiparticle prediction \eqref{MG41} is in perfect agreement with exact numerical calculations on the lattice, see Fig.\,\ref{Fig:MG4}. 
The time evolution of the Markov gap for $|C_{4,1}\ra$ is qualitatively identical to that for $|C_{3,1}\ra$. For $|C_{4,2}\ra$, the Markov gap has two local maxima, reflecting the different time windows of the $|\Q_{AB}|=2$ and $|\Q_{AB}|=3$ contributions.

Zero configurations affect the full tripartite entanglement dynamics, but most transparently its onset.  The species with maximal relative velocity belong to a zero-configuration. The next-fastest contributing configuration yields $t_e^{(4,2)}=\ell/\sqrt2$, instead of $t_e^{(4,1)}=\ell/2$. Thus, two quenches with the same propagation velocities can develop tripartite entanglement at different times because a crystal multiplet can contain uncorrelated quasiparticles.

\subsection{Long-time behavior}
\lb{longtime}

We investigate here the long-time behavior $t/\ell\to\infty$ of tripartite entanglement dynamics for general crystal quenches. As we shall see, this regime is governed by pairs of quasiparticles with same velocities.

\medskip

Suppose that three distinct quasiparticles $i,j,l$ belong to $A\cup B$. Their velocities must then satisfy
\bea\lb{3velineq}
|v_i(k)-v_j(k)|,\ |v_i(k)-v_l(k)|,\ |v_j(k)-v_l(k)|
\leq \frac{\ell_A+\ell_B}{t},
\eea
which in the limit $t/\ell\to\infty$ would impose $v_i(k)=v_j(k)=v_l(k)$. However, this cannot occur for three distinct velocities since a fixed value of the sine function has at most two distinct preimages modulo $2\pi$. Configurations with three or more quasiparticles in $A\cup B$ therefore cease to contribute at sufficiently large $t/\ell$.
The long-time behavior is thus governed by configurations of the form
\bea\lb{PairConfiguration}
\Q_A=\{a\},\qquad \Q_B=\{b\},\qquad \Q_C=\Q\setminus\{a,b\}\,.
\eea
With two quasiparticles in $A\cup B$, the remaining species necessarily lie in $C$ at sufficiently large $t/\ell$. The constraints on the emission point of the multiplet stemming from the remaining quasiparticles being in $C$ are therefore redundant in this late-time regime. 
For $a\in A$ and $b\in B$, the emission point $x\in \mathbb{R}$ of the multiplet must belong to the intersection
\bea
x\in[-\ell_A,0]\cap[-(v_b-v_a)t,\ell_B-(v_b-v_a)t],
\eea
where we have made the change of coordinate $x \rightarrow x+ v_a t$, and 
\bea
v_b-v_a = -2\sin(\pi (b-a) /p)\cos(k-\pi(a+b-2)/p)\,,
\eea
The kinematic weight $\D_{a,b}$ corresponding to this pair-configuration is given by
\bea\lb{LongTimeKernel}
\D_{a,b}(k,t)= \max\spm\big[0,\min\spm\big(0,\ell_B-(v_b -v_a) t\big)+\min\spm\big(\ell_A,(v_b -v_a) t\big)\big],
\eea
Introducing $n=b-a \;\,(\text{mod}\; p)$, and using the circulant form of the correlation matrix, the contribution of configuration \eqref{PairConfiguration} to the Markov gap depends on $n$ only,
\bea
m_{A:B}(a,b) = m_{A:B}(1,1+n) \equiv m_{A:B}(n)\,.
\eea
The late-time regime of the Markov gap quasiparticle picture becomes
\bea
M_{A:B}(t/\ell\gg1) &= \sum_{n=1}^{p-1}m_{A:B}(n)\sum_{a=1}^p\int_{\mathcal{B}}\frac{dk}{2\pi}\D_{a,a+n}(k,t)\,,\\
&=\sum_{n=1}^{p-1}m_{A:B}(n)\int_0^{2\pi}\hspace{-3pt}\frac{dk}{2\pi}\D_{1,1+n}(k,t)\,,
\eea
where in the second line we have translated the integration variable by $2\pi(a-1)/p$ to remove the dependence on $a$. The kinematic term depends on $k$ only through $v_{1+n}(k)-v_1(k) = -2\sin(\pi n /p)\cos(k-\pi n/p)$, where the phase and minus sign can be removed by another translation of $k$. Then for $x\in[-\ell_A,0]$, the condition that the second quasiparticle is in $B$ becomes
\bea
\frac{-x}{2t\sin(\pi n/p)} \le \cos k\le \frac{\ell_B-x}{2t\sin(\pi n/p)}\,,
\eea
while for sufficiently large $t/\ell$ we have $2t\sin(\pi n/p)>\ell_A+\ell_B$ such that both bounds lie between 0 and 1. We find
\bea
\frac1\ell\int_0^{2\pi}\hspace{-3pt}\frac{dk}{2\pi}\D_{1,1+n}(k,t)&=\frac1{\pi\ell}\int_{-\ell_A}^0\hspace{-2pt}dx \left[
\arcsin\spm\spm\left(\spm\frac{\ell_B-x}{2t\sin(\pi n/p)}\spm\right) - \arcsin\spm\spm\left(\spm\frac{-x}{2t\sin(\pi n/p)}\spm\right)\right]\spm,\\
&= \frac{r\ell}{2\pi t\sin(\pi n/p)} + O\!\left((\ell/t)^3\right)\spm,
\eea
and we finally obtain
\bea
\frac{M_{A:B}(t)}{\ell} = \frac\ell t \frac r{2\pi}\sum_{n=1}^{p-1}\frac{m_{A:B}(n)}{\sin(\pi n/p)}+ O\!\left((\ell/t)^3\right)\spm.
\eea
For $p>2$ the coefficient is positive, and the decay exponent is universal for infinite-systems.

The Markov gap density is straightforward to compute for arbitrary $(p,q)$,
\bea
m_{A:B}(n) &= 2s\Big(\frac12+\frac12\sqrt{\nu^2-g_n^2}+\frac12\sqrt{(1-\nu)^2-g_n^2}\Big)-2s(\nu) +s(\nu+g_n)+s(\nu-g_n)\,,\\
g_n&\equiv|G_{1,1+n}|=\frac{|\sin(\pi qn/p)|}{p\sin(\pi n/p)} \quad \text{and} \quad \nu=q/p.
\eea

\medskip

Zero-configurations affect the long-time behavior of the Markov gap, 
\bea
m_{A:B}(n)=0 \iff d\ \text{divides}\ n \iff g_n=0\,,
\eea
where we recall that $d=p/\gcd(p,q)$. Thus zero-configurations drop out of the long-time
coefficient. In particular, for even $p$ the term $n=p/2$ vanishes if and only if $\gcd(p,q)$ is even.

Note that these results only hold for finite $r$ in an infinite system, as the situation differs drastically for $r\rightarrow\infty$ or finite systems since any quasimomentum $k$ may contribute to tripartite entanglement at arbitrarily long time. For instance, if only $A$ is finite, the Markov gap grows monotonically and reaches a plateau at long times. In that case, for $|C_{p,1}\ra$ states we find 
\begin{equation}
    \frac{M^{(p,1)}_{A:B}(\infty)}{\ell_A}=\frac1p \sum_{|\mathcal{Q}_B|=1}^{p-2}m_{A:B}^{(p,1)}(|\mathcal{Q}_B|) \,.
\end{equation} 
where $m_{A:B}^{(p,1)}(|\mathcal{Q}_B|)$ is the Markov gap density \eqref{eq:cms-q1-gap} for $|\mathcal{Q}_A|=1$. 

\subsection{Finite systems: periodic and open boundary conditions}
\lb{finitesec}

We consider a partition into three adjacent intervals $A,B,C$ of a finite system length $L=\ell_A+\ell_B+\ell_C$, such that $A=[0,\ell_A),\,B=[\ell_A,\ell_A+\ell_B),\,C=[\ell_A+\ell_B,L)$.
For the $C_{3,1}$ quench, the three quasiparticle velocities are
$v_j(k)=\sin[k-2\pi(j-1)/3]$, with $k\in(\pi/3,\pi]$.
A triplet contributes to the Markov gap only when it has one species
in each interval. Its contribution is independent of momentum and
of the assignment of species to regions, and is given by $m^{(3,1)}_{A:B}$ in \eqref{mgdens31}.
We derive the full time dependence within the quasiparticle picture, first for periodic systems and then for reflecting open boundaries. Appendix \ref{App:FiniteMG} presents a different method than the one given here, applicable to any tripartite entanglement measure.

\subsubsection{Periodic boundary conditions}
Let $\chi_X$ be the indicator of $X=A,B,C$, extended periodically with period $L$. The six possible nontrivial quasiparticle configurations have equal integrals over a full momentum period. Extending the reduced
momentum interval to $[0,2\pi)$, we may thus write
\bea\lb{finite_gen}
 M_{\mathrm{PBC}}(t)
 =2m^{(3,1)}_{A:B}\int_0^L dx\int_0^{2\pi}\frac{dk}{2\pi}
 \chi_A(x+v_1t)\chi_B(x+v_2t)\chi_C(x+v_3t).
\eea
Each periodic indicator may be expanded as
\bea\lb{indicPBC}
 \chi_X(y)=\sum_{n\in\mathbb Z}c_{n}^X e^{2\pi\ii n y/L},
 \qquad
 c_{n}^X=\frac1L\int_X e^{-2\pi\ii n y/L}\,dy.
\eea
These coefficients are explicitly
\bea\lb{coefPBC}
 c_0^X=\frac{\ell_X}{L},\qquad
 c_{n\neq 0}^X=\frac{e^{-2\pi\ii n a_X/L}-e^{-2\pi inb_X/L}}{2\pi\ii n}.
\eea
where $X=[a_X,b_X)$. The integral over $x$ in \eqref{finite_gen} imposes $n_A+n_B+n_C=0$, yielding
\bea\lb{finitePBC}
 M_{\mathrm{PBC}}(t)
 =2m^{(3,1)}_{A:B} L\sum_{r,s\in\mathbb Z}c_r^A c_s^B c_{-r-s}^C\int_0^{2\pi}\frac{dk}{2\pi} \exp\!\left[\frac{2\pi\ii t}{L}
 \big(rv_1+sv_2-(r+s)v_3\big)\right]\spm.
\eea
Using the Bessel integral representation
$\int_0^{2\pi}\spm\spm\frac{dk}{2\pi}e^{i(a\sin k+b\cos k)}
=J_0(\sqrt{a^2+b^2})$, we obtain
\bea\lb{finitePBC-2}
 M_{\mathrm{PBC}}(t)
 =2m^{(3,1)}_{A:B} L\sum_{r,s\in\mathbb Z}c_r^A c_s^B c_{-r-s}^C
 J_0\!\left(\frac{2\pi t}{L}\sqrt{3(r^2+rs+s^2)}\right)\spm,
\eea
since $rv_1+sv_2-(r+s)v_3=\frac{3r}{2}\sin k-\frac{\sqrt3(r+2s)}{2}\cos k$. Terms with indices $(r,s)$ and $(-r,-s)$ are complex conjugates,
so the result is real.

\subsubsection{Open boundary conditions}
Reflections at $0$ and $L$ are implemented by the folding map $F(z)=|z|$ for $-L\le z\le L$, extended with period $2L$ (see Fig.\,\ref{Fig:TriEntOBC}). The physical position of quasiparticle $j$ is $F(x+v_jt)$. Thus $\widetilde\chi_R(z)=\chi_R(F(z))$ is an even, $2L$-periodic indicator, with expansion
\bea\label{finiteOBC}
 \widetilde\chi_X(z)=\sum_{n\in\mathbb Z}d_n^X e^{\ii\pi nz/L},
 \qquad
 d_n^X=\frac1L\int_X\cos\!\left(\frac{\pi ny}{L}\right)dy.
\eea
In particular,
\bea \label{coeffOBC}
 d_0^X=\frac{\ell_X}{L},\qquad
 d_{n \neq 0}^X=\frac{\sin(\pi nb_X/L)-\sin(\pi na_X/L)}{\pi n}.
\eea

We may proceed similarly to the periodic case, with $\chi_R$ replaced by $\widetilde\chi_R$. After the full momentum integral, the integrand is even in the position $x$: changing $x\mapsto-x$ can be compensated by $k\mapsto k+\pi$, because $F$ is even and $v_j(k+\pi)=-v_j(k)$. Consequently, the integral over the position $x$ can be replaced by one half of the integral over $[-L,L]$.
The same calculation as for the periodic case therefore yields
\bea\lb{finiteOBC2}
 M_{\mathrm{OBC}}(t)
 =2m^{(3,1)}_{A:B} L\sum_{r,s\in\mathbb Z}d_r^A d_s^B d_{-r-s}^C
 J_0\!\left(\frac{\pi t}{L}\sqrt{3(r^2+rs+s^2)}\right)\spm.
\eea
Reflection at the boundaries doubles the period and halves the Bessel frequencies. It also changes the Fourier coefficients, so the open and periodic results are generally not related by rescaling time alone.

\subsubsection{Late-time limit}

Since $r^2+rs+s^2=(r+s/2)^2+3s^2/4$, only the term $r=s=0$ has zero Bessel frequency. Every other term tends to zero as $t/L\to\infty$, and we obtain the same limit for both boundary conditions:
\bea
 M_\infty =2m^{(3,1)}_{A:B}\frac{\ell_A\ell_B\ell_C}{L^2}>0.
\eea

To describe the approach to the large time limit, let $P=L$ for periodic
boundaries and $P=2L$ for open boundaries. Group terms with the
same integer $n=r^2+rs+s^2$, and denote their coefficient sum by
\bea
 A_n=\sum_{\substack{r,s\in\mathbb Z\\r^2+rs+s^2=n}}
 \begin{cases}
 c_r^A c_s^B c_{-r-s}^C\spp,&\mathrm{PBC},\\[1ex]
 d_r^A d_s^B d_{-r-s}^C\spp,&\mathrm{OBC}.
 \end{cases}
\eea
Using $J_0(z)=\sqrt{\frac{2}{\pi z}}\cos\!\left(z-\frac\pi4\right)+O(z^{-3/2})$, gives, as $t/P\to\infty$,
\bea
 \frac{M(t)-M_\infty}L
 =2m^{(3,1)}_{A:B}\sqrt{\frac Pt}\sum_{n\ge1}\frac{A_n}{\pi(3n)^{1/4}}\cos\!\bigg(2\pi t\frac{\sqrt{3n}}P-\frac\pi4\bigg)+O\spm\spm \left(\spm L\left(t/P\right)^{-3/2}\right)\spm.
\eea
The quasiparticle prediction for the Markov gap thus relaxes with an oscillatory envelope of order $t^{-1/2}$. Each nonvanishing coefficient $A_n$ gives a frequency $\sqrt{3n}/P$.

The quasiparticle construction developed here for $|C_{3,1}\ra$ in systems with periodic and open boundary conditions extends to arbitrary crystal states $|C_{p,q}\ra$. As in infinite systems, zero configurations can delay the onset of the Markov gap and irreducible tripartite entanglement by eliminating contributing quasiparticle configurations. In finite geometries, they can also lower the nonzero late-time quasiparticle plateau of the Markov gap.

\section{Conclusion}
\label{Sec:Conclusion}

In this paper, we have investigated the multipartite entanglement structure and dynamics of the quasiparticle multiplets generated by crystal quenches in the tight-binding chain. Since the initial states are Gaussian, the entanglement carried by a multiplet is encoded in the correlation matrix of its quasiparticle species, while its spreading is set by their group velocities. 

Our central result is the identification of zero-configurations, which exist whenever $\gcd(p,q)>1$ and originate from the particular structure of the crystal correlation matrix. Although a multiplet in a zero-configuration is shared among three subsystems and entangled across every bipartition, its state consists entirely of bipartite-entangled states shared between pairs of subsystems, and thus carries no irreducible tripartite entanglement. We further showed that zero-configurations are the only nontrivial tripartitions of crystal multiplets with vanishing Markov gap, making the Markov gap a faithful witness of irreducible tripartite entanglement for crystal quenches. Besides zero-configurations, multiplets display a rich entanglement structure: $q=1$ multiplets are weighted W states, while larger multiplets with $q>1$ can host several independent tripartite resources, such as a product of two W states.

Zero-configurations have direct consequences on the dynamics. When both $p$ and $\gcd(p,q)$ are even, the pairs of quasiparticles with the largest relative velocity are uncorrelated, which further delays the onset of tripartite entanglement. This is most transparent for $|C_{4,1}\ra$ and $|C_{4,2}\ra$, whose quasiparticles propagate with identical velocities, yet tripartite entanglement starts building up at $t_e=\ell/2$ and $t_e=\ell/\sqrt2$. Zero-configurations also explain the nonpositive tripartite mutual information found after the quench from $|C_{4,2}\ra$ \cite{Caceffo:2023hns}, as every contributing quadruplet has a vanishing mutual information between two of the subsystems.

We have built the quasiparticle picture of the Markov gap. For $|C_{3,1}\ra$, it yields closed-form results that capture a delayed onset followed by a rapid growth, a second growth stage for subsystems of unequal sizes, and a nonmonotonic evolution characteristic of an entanglement barrier. For general crystal quenches in infinite systems, we showed that the late-time regime is governed by configurations with a single quasiparticle in each of the two finite subsystems, leading to a universal $1/t$ decay. For finite systems, we showed that the Markov gap remains nonzero at late times. Our predictions are in perfect agreement with exact lattice numerics in the hydrodynamic regime (see Appendix~\ref{Apdx:correl} for the time-dependent correlation functions used in numerical calculations).

Several directions remain open. Crystal multiplets shared among more than three parties display generalized zero-configurations, and a systematic study of their multipartite entanglement structure, using probes such as the genuine multientropy \cite{Gadde:2022cqi,Penington:2022dhr,Iizuka:2025caq,Iizuka:2025ioc,Harper:2024ker,Liu:2024ulq,Yuan:2024yfg,Berthiere:2025toi,Hu:2026bhg}, would be valuable. It would also be interesting to go beyond free fermions, for instance to interacting integrable models, and investigate whether GHZ-like entanglement can be produced after a quench.

\section*{Acknowledgments}
It is a pleasure to thank Gilles Parez for many interesting discussions and valuable comments on the draft.

\appendix
\renewcommand{\appendixname}{APPENDIX}
\counterwithin{equation}{section}
\renewcommand{\thesubsection}{\arabic{subsection}}

\section{Proof that the Markov gap vanishes only for zero-configurations}
\lb{vanishingMG}

The Markov gap can be equivalently viewed as a conditional mutual information
\bea
    M(A:B)&=S(AA^*)+S(AB)-S(A)-S(B)\,,\nn\\
    &=S(AA^*)+S(AB)-S(A)-S(ABA^*)\,,\nn\\
    &=I(A^*\spm:B|A)\,,\lb{MGalt1}
\eea
where the conditional mutual information is defined as
\bea
I(X:Z|Y) = I(X:Y\cup Z)-I(X:Y)\,.
\eea
Using purity of $ABA^*B^*$ and the symmetry $A\leftrightarrow B$ and $A,B\leftrightarrow A^*,B^*$ we have
\bea\lb{MGsym}
M(A:B)=I(A^*\spm:B|A)=I(B^*\spm:A|B)=I(A:B^*|A^*)=I(B:A^*|B^*)\,.
\eea
The reflected correlation matrix can be written in an more suggestive way as 
\bea\lb{GABABblock}
\mathbf{G}_{ABA^*B^*}=\begin{pmatrix}
\mathbf{G}_{aa} & \mathbf{G}_{ab} & \mathbf{G}_{aa^*} & \mathbf{G}_{ab^*} \\
\mathbf{G}_{ba} & \mathbf{G}_{bb} & \mathbf{G}_{ba^*} & \mathbf{G}_{bb^*}\\
\mathbf{G}_{a^*a} & \mathbf{G}_{a^*b} &\mathbf{G}_{a^*a^*} & \mathbf{G}_{a^*b^*}\\
\mathbf{G}_{b^*a} & \mathbf{G}_{b^*b} & \mathbf{G}_{b^*a^*} & \mathbf{G}_{b^*b^*}
\end{pmatrix}
=\begin{pmatrix}
G_{aa} & G_{ab} & D_{aa} & D_{ab} \\
G_{ba} & G_{bb} & D_{ba} & D_{bb}\\
D_{aa} & D_{ab} & 1-G_{aa} & -G_{ab}\\
D_{ba} & D_{bb} & -G_{ba} & 1-G_{bb}
\end{pmatrix}\spm,
\eea
where $D_{AB}\equiv\sqrt{G_{AB}(1-G_{AB})}$, and $D_{ab}$ is the restriction of $D_{AB}$ such that $a\in\Q_A$, $b\in\Q_B$, etc.

To proceed, we use an integral representation of the logarithm in terms of the resolvent,
\bea
\log G = -\int_0^\infty d\beta\bigg(\frac{1}{G+\beta}-\frac{1}{1+\beta} \bigg),
\eea
which gives the entropy $S$ for a correlation matrix $G$ as 
\bea
S=-\int_0^\infty d\beta\spp\beta\spp\Tr\bigg(\frac{1}{G+\beta}+\frac{1}{1-G+\beta} \bigg) +\int_0^\infty d\beta\spp\Tr\bigg(2-\frac1{1+\beta}\bigg)\spp.
\eea
Using the identity $\tr\big((G+\beta)^{-1}\big)=\partial_\beta \log\det(G+\beta)$ and integrating by parts yields
\bea
S\simeq\int_0^\infty d\beta\spp\log\det(G+\beta)\det(1-G+\beta),
\eea
up to non-universal terms that will cancel in the conditional mutual information. We obtain
\bea
I(X:Z|Y)=\int_0^\infty d\beta\spp\big( \Delta_\beta(\mathbf{G}) + \Delta_\beta(1-\mathbf{G})\big),\qquad \Delta_\beta(\mathbf{G}) \equiv \log\frac{\det(\mathbf{G}_{XY}+\beta)\det(\mathbf{G}_{YZ}+\beta)}{\det(\mathbf{G}_{Y}+\beta)\det(\mathbf{G}_{XYZ}+\beta)}\,.
\eea

Let us introduce the matrices
\bea
\begin{aligned}
K_{xx} &= \bG_{xx} +\beta - \bG_{xy}(\bG_{yy}+\beta)^{-1}\bG_{yx}\,, \\[-1ex]
K_{zz} &= \bG_{zz} +\beta - \bG_{zy}(\bG_{yy}+\beta)^{-1}\bG_{yz}\,, \qquad 
K_{XZ} = \begin{pmatrix}
K_{xx} & K_{xz} \\
K^\dagger_{xz} & K_{zz} 
\end{pmatrix}\spm. \\[-1ex]
K_{xz} &= \bG_{xz}  - \bG_{xy}(\bG_{yy}+\beta)^{-1}\bG_{yz}\,,
\end{aligned}
\eea
Using Schur's determinant identities
\bea
\begin{aligned}
&\det(\mathbf{G}_{XY}+\beta) = \det(\mathbf{G}_{yy}+\beta)\det K_{xx}\,,\qquad \quad\det(\mathbf{G}_{ZY}+\beta) = \det(\mathbf{G}_{yy}+\beta)\det K_{zz}\,,\\
&\det(\mathbf{G}_{XYZ}+\beta) = \det(\mathbf{G}_{yy}+\beta)\det K_{XZ}\,,\qquad\det K_{XZ}= \det K_{xx}\det(K_{zz}-K_{xz}^\dagger K_{xx}^{-1}K_{xz}) \,,
\end{aligned}
\eea
we obtain
\bea
\Delta_\beta(\mathbf{G})=\log\frac{\det K_{xx}\det K_{zz}}{\det K_{XZ}}\,.
\eea
Since $K_{XZ}$ is positive definite, Fischer's inequality gives $\Delta_\beta(\mathbf{G})\ge0$, with equality if and only if $K_{xz}=0$, and similarly for $1-\mathbf{G}$. A vanishing $I(X:Z|Y)$ thus requires $\Delta_\beta(\mathbf{G})=0=\Delta_\beta(1-\mathbf{G})$, i.e.
\bea
 \Delta_\beta(\mathbf{G})&=0 \iff \bG_{xz}  = \bG_{xy}(\bG_{yy}+\beta)^{-1}\bG_{yz}\,,\\
 \Delta_\beta(1-\mathbf{G})&=0 \iff \bG_{xz} = -\bG_{xy}(1-\bG_{yy}+\beta)^{-1}\bG_{yz}\,.
\eea
These two equations being valid for any positive $\beta$, letting $\beta\to\infty$ yields $\bG_{xz}=0$. 
For $\beta>0$, using Cayley-Hamilton theorem $(M+\beta)^{-1} = \sum_{n=0}^{m-1} \lambda_n(\beta)M^n$ where $m$ is the order of $M$, we get the constraints $\bG_{xy}\bG_{yy}^{n}\bG_{yz}=0$ for $n=0,1,\dots,|\Q_Y|-1$. We obtain that a vanishing conditional mutual information translates into
\bea\lb{IXZY0b}
I(X:Z|Y) = 0 \iff \bG_{xz}=0\;\; \text{and}\;\; \bG_{xy}\bG_{yy}^{n}\bG_{yz}=0 ,  \;\;n=0,1,\dots,|\Q_Y|-1 \,.
\eea

Specializing to the Markov gap \eqref{MGsym}, and using \eqref{GABABblock}, we have
\bea\lb{condn}
\begin{aligned}
&I(A^*\spm:B|A)= 0 \;\,\,\iff D_{ab}=0\;\; \text{and}\;\; D_{aa} G_{aa}^{n} G_{ab}=0 \,,\\
&I(B^*\spm:A|B)=0 \;\,\spp \iff D_{ba}=0\;\; \text{and}\;\; D_{bb} G_{bb}^{n} G_{ba}=0 \,,\\
&I(A:B^*|A^*)=0 \hspace{0.5pt}\iff D_{ab}=0\;\; \text{and}\;\; D_{aa} (1-G_{aa})^{n} G_{ab}=0 \,,\\
&I(B:A^*|B^*)=0 \iff D_{ba}=0\;\; \text{and}\;\; D_{bb} (1-G_{bb})^{n} G_{ba}=0 \,.
\end{aligned}
\eea
Now, using purity $G^2=G$, together with $D^2=G_{AB}(1-G_{AB})$, we get 
\bea
\begin{aligned}
&G_{xa}G_{ay}+G_{xb}G_{by}+G_{xc}G_{cy}=G_{xy}\\
&(D^2)_{ab} = 0 = G_{ab} - G_{aa}G_{ab}-G_{ab}G_{bb}\\
&(D^2)_{xx} = D^2_{xx}= G_{xx} - G_{xa}G_{ax}-G_{xb}G_{bx}
\end{aligned}
\quad\Rightarrow\quad
\begin{aligned}
&G_{ac}G_{cb}=0\\
&(D^2)_{xx} = G_{xc}G_{cx}
\end{aligned}
\eea
From \eqref{condn} with $n=0$ we have  
\bea
\begin{aligned}
&G_{xy} D_{yy}^2G_{yx} = 0\\
\Leftrightarrow\;& G_{xy} G_{yc}G_{cy}G_{yx} = 0\\
\Leftrightarrow\;& (G_{cy} G_{yx})^\dagger(G_{cy}G_{yx}) = 0\\
\Leftrightarrow\;& G_{xy} G_{yc} = 0
\end{aligned}
\eea
Finally, we arrive at a set of constraints on the correlation matrix $G$ only,
\bea\lb{cyclic}
M(A:B)=0 \;\;\iff\;\; G_{ab}G_{bc}=0\,, \quad G_{ca}G_{ab}=0\,, \quad  G_{bc}G_{ca}=0\,.
\eea
Note that these cyclic constraints are both necessary and sufficient for the Markov gap to vanish. Indeed, it is easy to check that they imply all the $n$ hierarchy in \eqref{IXZY0b}. Since $G$ is Hermitian, the three constraints read $G_{xy}G_{yz}=0$ for $(x,y,z)$ any cyclic permutation of $(a,b,c)$.
We next show that, for crystal multiplets, \eqref{cyclic} implies that at least one cross block $G_{xy}$ ($X\neq Y$) vanishes, i.e., that $\cP$ is a zero-configuration.
\medskip

Suppose that no cross block vanishes---we reason by contradiction.

\noindent Let $U_{jl}=p^{-1/2}e^{2\pi\ii lj/p}$ for $l=q+1,\dots,p$. Then $(V\;U)$ is the unitary Fourier matrix, such that
\bea
G_{xy}=V_XV_Y^\dagger=-U_XU_Y^\dagger\qquad (X\neq Y)\,,
\eea
with $\operatorname{ran}G_{yz}\subseteq\operatorname{ran}V_Y
\cap\operatorname{ran}U_Y$. Up to diagonal factors, $V_X$ and $U_X$ are Vandermonde matrices with
\bea
\operatorname{rank}V_X=\min(n_X,q)\,,\qquad \operatorname{rank}U_X=\min(n_X,p-q)\,,\qquad n_X=|\Q_X|\,.
\eea
Since $V_Y^\dagger$ and $U_Y^\dagger$ are both injective on $\operatorname{ran}G_{yz}$, applying $V_Y^\dagger$ or $U_Y^\dagger$ to $G_{yz}$ preserves rank, $\operatorname{rank}(V_Y^\dagger G_{yz})=\operatorname{rank}G_{yz}=\operatorname{rank}(U_Y^\dagger G_{yz})$. Writing $G_{xy}G_{yz}=V_X\big(V_Y^\dagger G_{yz}\big)=-U_X\big(U_Y^\dagger G_{yz}\big)=0$, Sylvester's rank inequality gives 
\bea\lb{bound1}
\rank\,G_{yz}\le q-\rank\, V_X\,, \qquad \rank\,G_{yz}\le p-q-\rank\,U_X\,.
\eea
Since we assumed that $G_{yz}\neq0$, this requires $n_X<\min(q,p-q)$ for every $X$, hence $\operatorname{rank}V_X=\operatorname{rank}U_X=n_X$.
Sylvester's inequality applied to $G_{yz}=V_YV_Z^\dagger=-U_YU_Z^\dagger$ also gives
\bea\lb{bound2}
\rank\,G_{yz}\ge n_Y+n_Z-q\,, \qquad\rank\, G_{yz}\ge n_Y+n_Z-(p-q)\,.
\eea
Combining bounds \eqref{bound1} and \eqref{bound2} yields $p\le2q\le p$, i.e., only half filling $q=p/2$ is permitted.

\smallskip 
For $p=2q$, the residue classes are the odd and even species: $G_{jk}=0$ for distinct $j,k$ of same parity, while $G_{jk}=\frac2p\frac{\omega^j}{\omega^k-\omega^j}$ for $j-k$ odd, with $\omega=e^{2\pi\ii/p}$. Up to a diagonal factor, any odd--even submatrix of $G$ is thus a Cauchy matrix, and has maximal rank.
Let $\Q_X^{\rm o}$ and $\Q_X^{\rm e}$ be the odd and even species of $\Q_X$, with $o_X=|\Q_X^{\rm o}|$ and $e_X=|\Q_X^{\rm e}|$. Since only odd--even entries of the cross blocks are nonzero, the entries of $G_{xy}G_{yz}$ between species of opposite parity vanish identically, and the constraint $G_{xy}G_{yz}=0$ reduces to
\bea\lb{eocycl}
\sum_{k\in\Q_Y^{\rm e}}G_{jk}\,G_{kl}=0\quad (j\in\Q_X^{\rm o},\ l\in\Q_Z^{\rm o})\,,\qquad
\sum_{k\in\Q_Y^{\rm o}}G_{jk}\,G_{kl}=0\quad (j\in\Q_X^{\rm e},\ l\in\Q_Z^{\rm e})\,.
\eea
Sylvester's inequality applied to \eqref{eocycl} gives 
\bea
\min(o_X,e_Y)+\min(e_Y,o_Z)\le e_Y\,, \qquad \min(e_X,o_Y)+\min(o_Y,e_Z)\le o_Y\,,
\eea
which if $o_X,o_Z,e_Y\ge1$ and $e_X,e_Z,o_Y\ge1$ require
\bea
o_X+o_Z\le e_Y\,, \qquad e_X+e_Z\le o_Y\,.
\eea
Since no two subsystems lie in the same parity class, one subsystem, say $Z$, contains both parities, and the two others can be labeled such that $o_X,e_Y\ge1$. With $Y$ and then $X$ in the middle of the cyclic constraint, we obtain
\bea
o_X+o_Z\le e_Y\qquad\text{and}\qquad e_Y+e_Z\le o_X\,.
\eea
Adding them gives $o_Z+e_Z\le 0$, which is a contradiction. Hence at least one cross block vanishes, i.e., $\cP$ is a zero-configuration.

Therefore, a vanishing Markov gap implies that at least one cross-correlation block vanishes. Together with \eqref{ZeroSufficiency}, this proves that the Markov gap density of a crystal multiplet vanishes if and only if the multiplet is in a zero-configuration.

\section{Markov gap evolution of $W$ states} 

\label{App:WStates}

\begin{figure}
    \centering
    \includegraphics[scale=1]{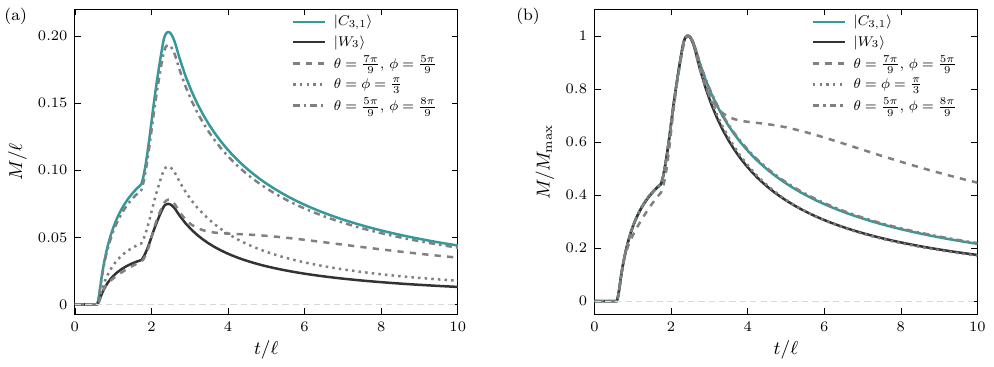}
    \caption{Markov gap time-evolution ($\ell_A=3\ell_B$) for $|C_{3,1}\rangle$, $|W_3\rangle$ and several $W$ states $|W(\theta,\phi)\rangle$. (a) Unnormalized curves. Different amplitudes of the Markov gap evolution are observed, depending on $\phi$ and $\theta$. (b) Normalized Markov gap $M/M_{\max}$. While most of the curves before $t_{max}$ are nearly identical, some values of $\phi \,, \theta$ show a different scaling of the Markov gap.}
    \label{Fig:WGen}
\end{figure}

In this section we review the Markov gap evolution of $|W_3\rangle$ states (see, e.g., \cite{Berthiere:2024sio}) defined as
\begin{equation}
|W_p\rangle = \frac{1}{p^{L/2p}} \prod_{i=1}^{L/p}\hspace{-1pt}\Bigg (\sum_{l=0}^{p-1} c_{pi-l}^\dagger \Bigg ) |0\rangle.
\label{Wstates}
\end{equation}
For such states, the quasiparticles dynamics is understandably identical to the one found for crystal states. However, its $G$ matrix reads
\begin{equation}
\begin{aligned}
G_{j,l}(k)&=\langle \psi_0|\eta_j^\dagger(k) \eta_l(k)|\psi_0\rangle \\
& = \frac{1}{L} \sum_{m,n=1}^L e^{-i(k-j2\pi/3)m+i(k-l2\pi/3)n} \langle c^\dagger_m c_n\rangle \\
& = \frac{1}{9} \sum_{m,n=1}^3 e^{-i(k-j2\pi/3)m+i(k-l2\pi/3)n} \,.
\end{aligned}
\end{equation}
The $k$-dependence of $G$ prevents a complete analytical study of the Markov gap density. The Markov gap evolution is thus written as
\begin{equation}
M^{(W_3)}_{A:B}(t)=\int_{\pi/3}^{\pi} \frac{dk}{2\pi} \sum_{Q_{AB}} m^{(W_3)}_{A:B}(k,Q_{AB})\big(\mathcal{D}_L(k,t)+\mathcal{D}_R(k,t) \big) \,.
\end{equation}
However, we can make a minor simplification. Indeed, the Markov gap density is identical for several partitions over certain ranges of $k$. The Markov gap evolution can thus be written in a more explicit way as
\begin{equation}
\begin{aligned}
M^{(W_3)}_{A:B}(t)&=\int_{\pi/3}^{\pi/2} \frac{dk}{2\pi} \Big(m^{(W_3)}_{A:B}(k,\{\{2\},\{0\}\})\mathcal{D}_L(k,t) + m^{(W_3)}_{A:B}(k,\{\{1\},\{2\}\})\mathcal{D}_R(k,t) \Big) \\
&\hspace{1cm}+\int_{\pi/2}^{5\pi/6} \frac{dk}{2\pi} \Big(m^{(W_3)}_{A:B}(k,\{\{1\},\{0\}\})\mathcal{D}_L(k,t) + m^{(W_3)}_{A:B}(k,\{\{2\},\{1\}\})\mathcal{D}_R(k,t) \Big) \\
&\hspace{2cm}+\int_{5\pi/6}^{\pi} \frac{dk}{2\pi} \Big(m^{(W_3)}_{A:B}(k,\{\{0\},\{1\}\})\mathcal{D}_L(k,t) + m^{(W_3)}_{A:B}(k,\{\{2\},\{0\}\})\mathcal{D}_R(k,t) \Big) \,,
\end{aligned}
\end{equation}
We compare results from $|W_3\rangle$ and $|C_{3,1}\rangle$ states in Fig.$\,$\ref{Fig:WGen}. While the overall entanglement is always higher for $|C_{3,1}\rangle$, the normalized entanglement shows that the curves are identical until they reach their peak, after which they separate into different decreasing slopes. This behavior raised the question of whether the initial slope is universal. To study this further, we introduce the generalized $W$ states, defined as
\begin{equation}
|W(\theta,\phi)\rangle= \big(\cos \theta |001\rangle+\sin \theta \cos \phi |010 \rangle + \sin \theta \sin \phi |100 \rangle\big)^{\otimes L/3} \,,
\end{equation}
and compare it to $|W_3\rangle$ and $|C_{3,1}\rangle$ for different values of $\theta$ and $\phi$ in Fig.$\,$\ref{Fig:WGen}. We observe that pre-max curves are similar but not identical, and that $|W_3\rangle$ is not the state of least entanglement in general. Importantly, for $\theta=7\pi/9$ and $\phi=5\pi/9$, the generalized $W$ state seems to decay in a very unique way after its peak. This behavior is not observed in other states. Our results are in perfect agreement with numerical simulations in the hydrodynamic regime \mbox{(see \cite{Berthiere:2024sio}).}

In conclusion, although all studied states exhibit similar behavior, particularly before reaching their peak, no universal slope appears across configurations. Furthermore, it would be interesting to know more about why particular configurations of the generalized $W$ state exhibit such a unique decay.

\medskip

Finally, similarly to what was observed for crystal states, taking $\ell_A \rightarrow \infty$ makes the Markov gap behave analogously to bipartite entropy. In particular, it reaches a stationary value at infinity given by
\begin{equation}
M_{A:B}^{(W_3)}(t\rightarrow \infty)=\ell_B \int_{\pi/3}^\pi \frac{dk}{2\pi} \sum_{\mathcal{Q}_{A}}m^{(W_3)}_{A:B}(k,\Q_A,\Q_B=\{M\}) \,,
\end{equation}
in a very similar fashion to the bipartite prediction.

\section{Phenomenology of the quasiparticles triplets}

\label{App:Pheno}

We study here in detail the kinematic weights $\mathcal{D}_L$ and $\mathcal{D}_R$ beyond the semi-analytic description. While \eqref{D1MM} and \eqref{D2MM} are sufficient to evaluate the Markov gap through the semi-analytic formula \eqref{MarkovGap}, a more in-depth study of the quasiparticles phenomenology allows the derivation of a closed form (evaluated in Appendix\,\ref{App:ExactDerivation}) and provides useful insight into the behavior of tripartite entanglement following a quench.

\medskip

As seen previously, the triplets originating from the region $\mathcal{D}_{L/R}$ all provide entanglement at time $t$ and momentum $k$. In the following, we consider the boundary points of these regions as moving points, and study their evolution through time. For both assignments, we consider the setup depicted in Fig.\,\ref{Fig:BehaviorD1}(a), with $x=0$ being the boundary between $A$ and $B$. Importantly, we assume $\ell_A \geq\ell_B$. This is done without loss of generality because of $M(A:\spm B)=M(B:\spm A)$ and symmetry with respect to $v\leftrightarrow-v$.

We restrict the quasiparticles distribution to $k\in [\pi/3,2\pi/3]$ for which $v_M(k)<0$ for the moment. We therefore label the obtained contributions $\mathcal{D}_R^{(-)}$ and $\mathcal{D}_L^{(-)}$. Similarly, $\mathcal{D}_R^{(+)}$ and $\mathcal{D}_L^{(+)}$, corresponding to $k\in [2\pi/3,\pi]$, are derived analogously and quickly reviewed at the end of the section.

\subsection{Evaluation of $\mathcal{D}_L^{(-)}$}

\label{App:Pheno1}

\begin{figure}[b]
\centering
\hspace{-0.5cm}
\begin{minipage}{.5\textwidth}
  \centering
\includegraphics[scale=1.4]{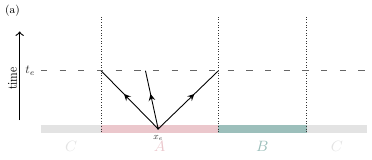}
\vspace{0.1cm}
\end{minipage}%
\begin{minipage}{.5\textwidth}
  \centering
\includegraphics[scale=1.4]{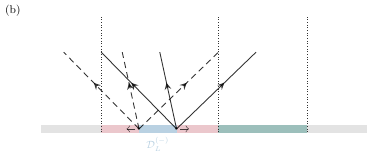}
\vspace{0.1cm}
\end{minipage}
\hspace{-0.5cm}
\begin{minipage}{.5\textwidth}
\hspace{-0.6cm}
  \centering
\includegraphics[scale=1.4]{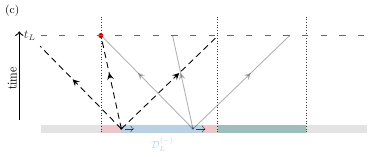}
\end{minipage}%
\begin{minipage}{.5\textwidth}
\hspace{-0.6cm}
  \centering
\includegraphics[scale=1.4]{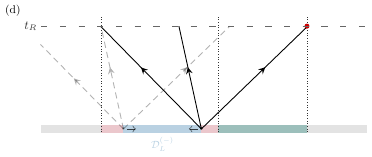}
\end{minipage}
\vspace{-5pt}
\caption{Behavior of entanglement fronts throughout the evolution of the system for $\mathcal{D}^{(-)}_L$ and a choice of $k$. (a) At $t=t_e$, all the quasiparticles have just reached their assigned systems. (b) The system decouples into two fronts propagating away from each other, such that all the quasiparticles stay in their assigned subregions. Entanglement increases. (c) The middle quasiparticle of the left front reaches the boundary between $B$ and $C$, forcing $x_L$ to counter-propagate to meet entanglement conditions. (d) The same phenomenon happens at the right front for the rightmost quasiparticle, making both fronts propagate toward each other and decreasing entanglement generation. They eventually meet at $t=t_F$.}
\label{Fig:BehaviorD1}
\end{figure}

In this situation, the rightmost quasiparticle with velocity $v_R$ is assigned to the region $B$, and the remaining quasiparticles with velocities $v_M$ and $v_L$ respectively go to $A$ and the left side of $C$.
We derive the initial position $x_e$ that provides entanglement the earliest. Understandably, $x_e$ will be located in $A$, as QP $L$ and $R$ will both reach their assigned regions the fastest this way. The time at which each of them crosses the boundaries of $A$ is
\begin{equation}
t_1(x)=\frac{-x}{v_R(k)} \,, \quad t_2(x)=-\frac{x+\ell_A}{v_L(k)} \,.
\label{InitTime}
\end{equation}
We then derive the initial position by setting $t_1=t_2$ and solving for $x$. Solving in turn \eqref{InitTime} for $t$ yields the onset time $t_e$. They read
\begin{equation}
x_e=\frac{-\ell_A v_R(k)}{v_R(k)-v_L(k)} \,, \quad t_e=\frac{\ell_A}{v_R(k)-v_L(k)} \,.
\end{equation}
 Fig.\,\ref{Fig:BehaviorD1}(a) presents the situation at $t=t_e$. We now split $x$ into $x_R$ and $x_L$, the boundaries of $\mathcal{D}_L^{(-)}$ spreading in opposite directions. At $t>t_e$, they begin to spread at the maximum velocity such that each quasiparticle is maintained in its respective region, i.e. $x_R$ propagates at $-v_L$ and $x_L$ at $-v_R$ (see Fig.\,\ref{Fig:BehaviorD1}(b)). They are then defined respectively as
\begin{equation}
\begin{aligned}
&x_R(t)=x_e-v_{L}(k)(t-t_{e})=-\ell_{A}-v_{L}(k)t \,, \\
&x_L(t)=x_e-v_{R}(k)(t-t_{e})=-v_{R}(k)t \,,
\end{aligned}
\end{equation}
where we naturally imposed continuity with $x_e$ at $t_e$.

At $t=t_L$, the middle quasiparticle propagating from $x_L$ reaches $-\ell_A$. This forces the left front to counter-propagate for entanglement to remain. This is pictured in Fig.\,\ref{Fig:BehaviorD1}(c). The same phenomenon is observed for $x_R$ at $t_R$ and is shown Fig.\,\ref{Fig:BehaviorD1}(d). The corresponding formulas for $t_R$ and $t_L$ are
\begin{equation}
\begin{aligned}
&x_R(t_R)+v_{R}(k)t_R=\ell_{B} \iff t_R=\frac{\ell_{A}+\ell_{B}}{v_{R}(k)-v_{L}(k)} \,,\\
&x_L(t_L)+v_{M}(k)t_L=-\ell_{A} \iff t_L=\frac{\ell_{A}}{v_{R}(k)-v_{M}(k)} \,,
\end{aligned}
\end{equation} 
and $x_{L/R}(t>t_{L/R})$ are obtained as
\begin{equation}
\begin{aligned}
&x_R(t>t_R)=\ell_{B}-v_{R}(k)t \,,\\
&x_L(t>t_L)=-\ell_{A}-v_{M}(k)t \,.\\
\end{aligned}
\end{equation}
After that, both fronts propagate towards each other and meet at $t_F$, defined as
\begin{equation}
x_R(t_F)-x_L(t_F)=0 \iff t_F=\frac{\ell_{A}+\ell_{B}}{v_{R}(k)-v_{M}(k)} \,.
\end{equation}
The entanglement evolution $\mathcal{D}^{(-)}_L=x_R(t)-x_L(t)$ then reads
\begin{equation}
\mathcal{D}_L^{(-)}(k,\ell,t)=\left \{ \begin{matrix*}[l]
0 \,,&\quad t<t_{e} \lor t\geq t_{F} \,, \\
\big (v_{R}(k)-v_{L}(k) \big)t-\ell_{A} \,, &\quad t_{e}\leq t<\min(t_R,t_L)\,, \\
\big(v_{M}(k)-v_{L}(k)\big)t \,, &\quad t_L\leq t<t_R\,, \\
\ell_{B} \,, &\quad t_R\leq t<t_L\,, \\
\ell_{A}+\ell_{B} -\big(v_{R}(k)-v_{M}(k)\big)t \,, &\quad \max(t_R,t_L)\leq t<t_F \,, \\
\end{matrix*} \right .
\label{piecewise1}
\end{equation}
Since each behavior is worked out to maximize entanglement, the total quantity can be computed by taking the minimal value among every individual behavior and written as
\begin{equation}
\mathcal{D}_L^{(-)}(k,\ell,t)=\max \bigg [ \min \Big (\big(v_{R}(k)-v_{L}(k)\big)t-\ell_{A},\big(v_{M}(k)-v_{L}(k)\big)t,\ell_{B},\ell_{A}+\ell_{B}-\big(v_{R}(k)-v_{M}(k)\big)t \Big),0\bigg ] \,,
\end{equation} 
which is equal to \eqref{D1MM} after taking the appropriate simplifications due to $\ell_A \geq \ell_B$ and $v_M<0$, as expected.

\subsection{Evaluation of $\mathcal{D}_R^{(-)}$}

\label{App:Pheno2}

\begin{figure}[b]
\centering
\hspace{-0.5cm}
\begin{minipage}{.5\textwidth}
  \centering
\includegraphics[scale=1.4]{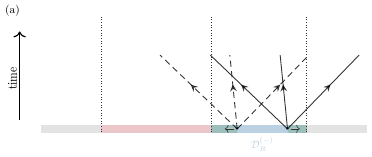}
\vspace{0.5cm}
\end{minipage}%
\begin{minipage}{.5\textwidth}
  \centering
\includegraphics[scale=1.4]{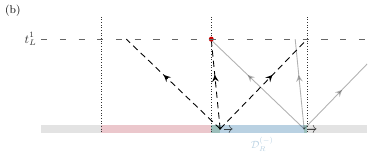}
\vspace{0.5cm}
\end{minipage}
\hspace{-0.5cm}
\begin{minipage}{.5\textwidth}
\hspace{-0.6cm}
  \centering
\includegraphics[scale=1.4]{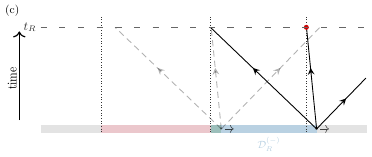}
\end{minipage}%
\begin{minipage}{.5\textwidth}
\hspace{-0.6cm}
  \centering
\includegraphics[scale=1.4]{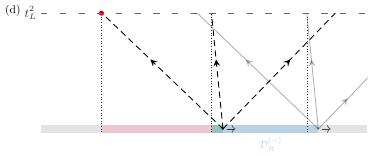}
\end{minipage}
\vspace{-5pt}
\caption{Behavior of entanglement fronts throughout the evolution of the system for $\mathcal{D}^{(-)}_R$ and a choice of $k$. (a) Entanglement spreading after $t_e$. Both fronts propagate away from each other, increasing entanglement. (b) The middle quasiparticle of the left front reaches the boundary between $A$ and $B$. $x_L$ is forced to counterpropagate to keep each quasiparticle in its assigned subregion. (c) The middle quasiparticle of the right front reaches the boundary between $B$ and $C$. This forces $x_R$ to slow down its propagation to $-v_M$. (d) The left quasiparticle of the left front reaches the boundary of $A$. $x_L$ thus increases its velocity to $-v_R$. Since by definition $-v_L\geq -v_M$, $x_L$ will chase $x_R$ until they eventually meet at $t=t_F$. Single and double arrows respectively stand for slow and rapid movement.}
\label{Fig:BehaviorD2}
\end{figure}

The second assignment that produces entanglement consists of assigning the leftmost quasiparticle to $A$, and the two remaining to $B$ and the rightmost side of $C$, as represented in Fig.\,\ref{Fig:BehaviorD2}. The derivation is similar. This time, $x_e$ will be located inside of $B$, with 
\begin{equation}
t_1(x)=\frac{\ell_B-x}{v_R(k)} \,, \quad t_2(x)=-\frac{x}{v_L(k)} \,.
\end{equation}
The initial position and time follows as
\begin{equation}
x_e=\frac{-\ell_B v_L(k)}{v_R(k)-v_L(k)} \,, \quad t_e =\frac{\ell_B}{v_R(k)-v_L(k)} \,,
\end{equation}
The post-$t_e$ fronts of entanglement are derived as
\begin{equation}
\begin{aligned}
&x_R(t)=x_e-v_{L}(k)(t-t_{e})=-v_{L}(k)t \,, \\
&x_L(t)=x_e-v_{R}(k)(t-t_{e})=\ell_{B}-v_{R}(k)t \,.
\end{aligned}
\end{equation}
From here, the left front will first counter-propagate because of the middle quasiparticle reaching the left boundary of $B$ at $t=t_L^1$ (see Fig.\,\ref{Fig:BehaviorD2}(b)). Its velocity is then changed a second time upon $x_L-v_Lt$ reaching the left boundary of $A$ at time $t=t_L^2$ (see Fig.\,\ref{Fig:BehaviorD2}(d)). The right front behaves similarly in that it will be forced into a slower propagation at time $t=t_R$ (see Fig.\,\ref{Fig:BehaviorD2}(c)).
The boundary times are derived as
\begin{equation}
\begin{aligned}
&x_R(t_R)+v_{M}(k)t_R=\ell_{B} \iff t_R=\frac{\ell_{B}}{v_{M}(k)-v_{L}(k)} \,,\\
&x_L(t_L^1)+v_{M}(k)t_L^1=0 \iff t_{L}^1=\frac{\ell_{B}}{v_{R}(k)-v_{M}(k)} \,.\\
&x_L(t_L^2)+v_{L}(k)t_L^2=-\ell_{A} \iff t_{L}^2=\frac{\ell_{A}}{v_{M}(k)-v_{L}(k)} \,.\\
\end{aligned}
\end{equation}
Notice that $t_R\leq t_L^2$. The entanglement fronts $x_{R/L}$ after those times are
\begin{equation}
\begin{aligned}
&x_R(t>t_R)=\ell_{B}-v_{M}(k)t \,,\\
&x_{L}(t_L^2\geq t>t_{L}^1)=-v_{M}(k)t \,,\\
&x_{L}(t>t_{L}^2)=-\ell_{A}-v_{L}(k)t \,.\\
\end{aligned}
\end{equation}
Because the two fronts do not counterpropagate toward each other but the left front rather "chases" the right one at a higher velocity, $\mathcal{D}^{(-)}_R$ decays more slowly. For a well-chosen $k$, it only vanishes at $t\rightarrow\infty$. Both fronts meet at $t_F$, defined as
\begin{equation}
x_R(t_F)-x_L(t_F)=0\iff t_F=\frac{\ell_{A}+\ell_{B}}{v_{M}(k)-v_{L}(k)} \,.
\end{equation} 
Finally, $\mathcal{D}^{(-)}_R(k,t)$ reads
\begin{equation}
\mathcal{D}^{(-)}_R(k,t)=\left \{ \begin{matrix*}[l]
	0\,, &\quad t<t_e\lor t\geq t_F \,, \\
\big (v_{R}(k)-v_{L}(k) \big)t -\ell_{B}\,, &\quad t_e\leq t< t_L^1 \,, \\
\big (v_{M}(k)-v_{L}(k) \big)t \,, &\quad t_{L}^1\leq t<t_R \,, \\
\ell_{B}\,, &\quad  t_{R}\leq t<t_{L}^2\,, \\
\big (v_{L}(k)-v_{M}(k) \big)t+\ell_{A}+\ell_{B}\,, &\quad t_L^2\leq t<t_F \,. \\
\end{matrix*} \right .
\label{piecewise2}
\end{equation}
As before, \eqref{piecewise2} can be rewritten as
\begin{equation}
\mathcal{D}^{(-)}_R(k,t)=\max \bigg [ \min \Big (\big(v_{R}(k)-v_{L}(k)\big)t-\ell_{B},\big(v_{M}(k)-v_{L}(k)\big)t,\ell_{B},\big(v_{L}(k)-v_{M}(k)\big)t+\ell_{A}+\ell_{B} \Big),0\bigg ] ,
\end{equation} 
which coincides with \eqref{D2MM} after simplifications.

\subsection{Evaluation of $\mathcal{D}_L^{(+)}$ and $\mathcal{D}_R^{(+)}$}

Contributions for $k$ such that $v_M(k)>0$ are obtained by symmetry of the system. Indeed, by taking the mirror image of the system, we revert back to $v_M(k)<0$, with $B$ to the left of $A$, but still with $k\in[\pi/3,2\pi/3]$ ($k\in [2\pi/3,\pi]$ can be recovered by taking $(v_R,v_M,v_L) \rightarrow -(v_L,v_M,v_R)$ in the following results). The same phenomenological study can be performed considering this time $A$ of length $\ell_B$ and $B$ of length $\ell_A$, still with $\ell_A \geq \ell_B$. $\mathcal{D}_L^{(+)}$ and $\mathcal{D}_R^{(+)}$ read
\begin{equation}
\mathcal{D}_L^{(+)}(k,t)=\left \{ \begin{matrix*}[l]
	0\,, &\quad t<t_e\lor t\geq t_F \,, \\
\big (v_{R}(k)-v_{L}(k) \big)t -\ell_{A}\,, &\quad t_e\leq t<\min(t_L^1,t_L^2) \,, \\
\big (v_{M}(k)-v_{L}(k) \big)t \,, &\quad t_{L}^1\leq t<t_L^3 \,, \\
\ell_{B}\,, &\quad  \max(t_{L}^2,t_L^3)\leq t<t_{R}\,, \\
\big (v_{L}(k)-v_{M}(k) \big)t+\ell_{A}+\ell_{B}\,, & \quad t_R\leq t<t_F \,, \\
\end{matrix*} \right .
\label{piecewise3}
\end{equation}
with
\begin{equation}
\begin{aligned}
t_e=\frac{\ell_A}{v_R(k)-v_L(k)} \,, \qquad t_L^1=\frac{\ell_A}{v_R(k)-v_M(k)} \,, \\
t_L^2=\frac{\ell_A+\ell_B}{v_R(k)-v_L(k)} \,, \qquad t_L^3=\frac{\ell_B}{v_M(k)-v_L(k)} \,, \\
t_R= \frac{\ell_A}{v_M(k)-v_L(k)} \,, \qquad t_F=\frac{\ell_A+\ell_B}{v_M(k)-v_L(k)} \,,
\end{aligned}
\end{equation}
and
\begin{equation}
\mathcal{D}_R^{(+)}(k,t)=\left \{ \begin{matrix*}[l]
0 \,,&\quad t<t_{e} \lor t\geq t_{F} \,, \\
\big (v_{R}(k)-v_{L}(k) \big)t-\ell_{B} \,, &\quad t_{e}\leq t<t_L\,, \\
\big(v_{M}(k)-v_{L}(k)\big)t \,, &\quad t_L\leq t<\min(t_R^1,t_R^2)\,, \\
\ell_{B} \,, &\quad t_R^2\leq t<t_R^3\,, \\
\big(v_{M}(k)-v_{R}(k)\big)t+\ell_{A}+\ell_{B} \,, &\quad \max(t_R^1,t_R^3)\leq t<t_F \,, \\
\end{matrix*} \right .
\label{piecewise4}
\end{equation}
with
\begin{equation}
\begin{aligned}
t_e=\frac{\ell_B}{v_R(k)-v_L(k)} \,, \qquad t_R^1=\frac{\ell_A+\ell_B}{v_R(k)-v_L(k)} \,, \\
t_R^2=\frac{\ell_B}{v_M(k)-v_L(k)} \,, \qquad t_R^3=\frac{\ell_A}{v_R(k)-v_M(k)} \,, \\
t_L= \frac{\ell_B}{v_R(k)-v_M(k)} \,, \qquad t_F=\frac{\ell_A+\ell_B}{v_R(k)-v_M(k)} \,.
\end{aligned}
\end{equation}
They can also be written in terms of min and max functions and read
\begin{equation}
    \mathcal{D}_L^{(+)}=\max\bigg[ \min \Big( \big(v_R(k)-v_L(k)\big)t-\ell_A, \big(v_M(k)-v_L(k)\big)t,\ell_B,\big(v_L(k)-v_M(k)\big)t+\ell_A+\ell_B\Big),0\bigg]\,,
\end{equation}
and
\begin{equation}
    \mathcal{D}_R^{(+)}=\max\bigg[ \min \Big( \big(v_R(k)-v_L(k)\big)t-\ell_B, \big(v_M(k)-v_L(k)\big)t,\ell_B,\big(v_M(k)-v_R(k)\big)t+\ell_A+\ell_B\Big),0\bigg] \,.
\end{equation}

\section{Closed-form evaluation of (\ref{MarkovGap})}
\label{App:ExactDerivation}

\begin{figure}[b]
  \centering
\includegraphics[scale=1.0]{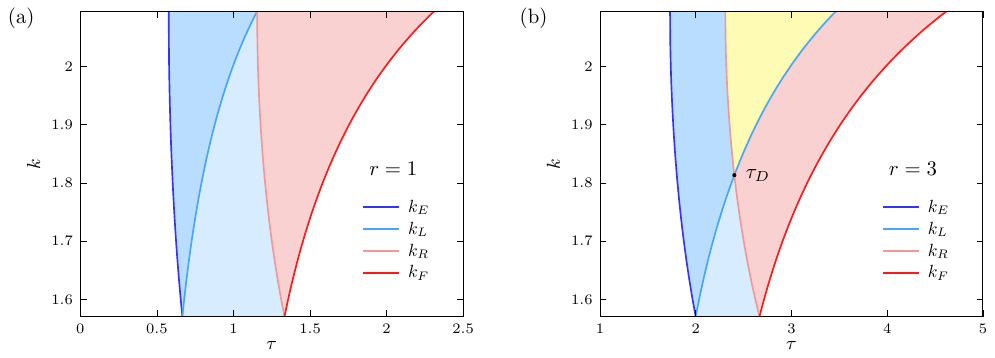}
\caption{Integration regions. Regions inside which entanglement is increasing are denoted in blue (the darker the shade, the faster the entanglement rises), while decreasing regions are denoted in red. The yellow region represents a constant entanglement. (a) $\ell_A=\ell_B$. (b) $\ell_A=3\ell_B$. Here $\tau_D$ represents a key time after which the entanglement stops solely increasing and starts decaying for some momenta.}
\label{Fig:MomentumBehaviorA}
\end{figure}

In this section, we build on the previous section to analytically derive the Markov gap evolution following a quench of $|C_{3,1}\rangle$ crystal states. Let us begin by rewriting \eqref{MarkovGap} as
\begin{equation}
\begin{aligned}
M_{A:B}(t)&=m_{A:B}^{(3,1)} \int_{\pi/3}^\pi \frac{dk}{2\pi} \big(\mathcal{D}_L(k,\ell_A,\ell_B,t)+\mathcal{D}_R(k,\ell_A,\ell_B,t)\big)\\
&=m_{A:B}^{(3,1)}\, \ell_{B} \int_{\pi/3}^\pi \frac{dk}{2\pi} \big( \mathcal{D}_L(k,r,1,t/\ell_{B})+\mathcal{D}_R(k,r,1,t/\ell_{B})\big)\\
&=\frac{m_{A:B}^{(3,1)} \,\ell_{B}}{2\pi} \big(I_L(\tau,r)+I_R(\tau,r)\big) \,,
\end{aligned}
\end{equation}
where $r=\ell_A/\ell_B$ and $\tau=t/\ell_B$. Focusing on $I_L(\tau,r)$ for the moment, we further break it down to
\begin{equation}
I_L^{(-)}=\int_{\pi/2}^{2\pi/3} dk \, \mathcal{D}^{(-)}_L(k,r,\tau) \,,
\end{equation}
where we have arbitrarily chosen a subregion of $[\pi/3,\pi]$ in which $v_R,v_M$ and $v_L$ correspond to constant quasiparticle indices. Within this region, the velocities obey the following relationships
\begin{equation}
\begin{aligned}
& v_R(k)-v_L(k)= \sqrt{3}\sin \Big (k -\frac{\pi}{6} \Big) \,, \\
& v_R(k)-v_M(k)=\sqrt{3} \sin \Big (k+\frac{\pi}{6} \Big) \,,\\
&  v_M(k) - v_L(k)=-\sqrt{3} \cos(k) \,.
\end{aligned}
\end{equation}
To integrate $\mathcal{D}^{(-)}_L$ with respect to $k$, we must work out the values of $k$ corresponding to a change in behavior. We thus invert the boundary $t$ equations found in Appendix\,\ref{App:Pheno1} and find
\begin{equation}
\begin{aligned}
t_{e}=\frac{\ell_{A}}{\sqrt{3}\sin(k-\pi/6)} &\iff k_{e}=\frac{\pi}{6}+ \arcsin\bigg(\frac{r}{\sqrt{3}\tau}\bigg) \,,\\
t_R=\frac{\ell_{A} + \ell_{B}}{\sqrt{3}\sin(k-\pi/6)} &\iff k_R=\frac{\pi}{6}+ \arcsin\bigg(\frac{r+1}{\sqrt{3}\tau}\bigg)\,, \\
t_L=\frac{\ell_{A}}{\sqrt{3}\sin(k+\pi/6)} &\iff k_L=\frac{5\pi}{6}- \arcsin\bigg(\frac{r}{\sqrt{3}\tau} \bigg)\,, \\
t_{F}=\frac{\ell_{A}+\ell_{B}}{\sqrt{3}\sin(k+\pi/6)} &\iff k_{F}=\frac{5\pi}{6}- \arcsin\bigg(\frac{r+1}{\sqrt{3}\tau} \bigg)\,, \\
\end{aligned}
\end{equation}
where we have chosen each branch such that they have non-zero support in $[\frac{\pi}{2},\frac{2\pi}{3}]$.$\spm$ The integration of $\mathcal{D}^{(-)}_L \spm$ then reads
\begin{equation}
I^{(-)}_L(\tau,r)=\sum_{i=1}^{N(\tau)-1} \int_{k_i}^{k_{i+1}} dk\,\mathcal{D}^{(-)}_L(k,r,\tau) \,,
\end{equation} 
with $k_1=\pi/2$, $k_{N(\tau)}=2\pi/3$ and $k_1<k_2<\hdots<k_{N(\tau)}$. $N(\tau)$ varies according to the number of $k_i$ encountered at time $\tau$. To get a clearer picture, we plot $k$ vs $\tau$ for different values of $r$ in Fig.$\,$\ref{Fig:MomentumBehaviorA}. We integrate for each $\tau$ a vertical slice of the graphs, and sum the contribution of each colored region encountered. Fig.$\,$\ref{Fig:MomentumBehaviorA} gives insights as to how the final result will behave. In particular, we observe a crossing at $\tau_D=2\sqrt{1+r+r^2}/3$ between a sector of entanglement growth and one of entanglement decay. This shows how entanglement will decrease soon after this point due to a larger portion of momentum being in the red or yellow zone. 

Using \eqref{piecewise1} and the analysis above, we finally derive each different behavior and find
\begin{equation}
\scalemath{0.89}{
I^{(-)}_L(\tau,r)= \left \{\begin{matrix*}[l]
0 \,, & \tau < \frac{r}{\sqrt{3}} \lor \frac{2(r+1)}{\sqrt{3}}\leq\tau  \,, \\[10pt]
\sqrt{3\tau^2-r^2}-r \arccos \frac{r}{\sqrt{3}\tau} \,, & \frac{r}{\sqrt{3}} \leq \tau < \min \Big(\frac{2r}{3},\frac{1+r}{\sqrt{3}}\Big)\,, \\[10pt]

\hspace{-2.7cm}\begin{matrix}
\frac{\pi}{2}+\sqrt{3\tau^2-r^2}-\sqrt{3\tau^2-(r+1)^2} \\[4pt]
\hspace{5.5cm} +r \arcsin \frac{r}{\sqrt{3}\tau}-(r+1) \arcsin \frac{r+1}{\sqrt{3}\tau} \,,
\end{matrix}
 & \frac{1+r}{\sqrt{3}} \leq \tau < \frac{2r}{3} \text{ if } r>3+2\sqrt{3}\,, \\[18pt]

-\frac{r\pi}{3}+\sqrt{3}\tau-\sqrt{3\tau^2-r^2} +r\arccos \frac{r}{\sqrt{3}\tau}\,, & \frac{2r}{3} \leq \tau < \frac{1+r}{\sqrt{3}} \text{ else} \,, \\[10pt]

\hspace{-1.5cm}\begin{matrix}
-\frac{r\pi}{3}+\sqrt{3}\tau-\sqrt{3\tau^2-r^2}-\sqrt{3\tau^2-(1+r)^2} \\[4pt] \hspace{4.3cm}+(1+r)\arccos \frac{r+1}{\sqrt{3}\tau}+r\arccos \frac{r}{\sqrt{3}\tau}\,,
\end{matrix}

 & \max \Big(\frac{2r}{3},\frac{1+r}{\sqrt{3}} \Big) \leq \tau < \min\Big(\frac{2(r+1)}{3},\frac{2r}{\sqrt{3}}\Big) \,, \\[18pt]
\hspace{-2.7cm}\begin{matrix}
-\frac{\pi}{6}+\sqrt{3\tau^2-(1+r)^2}-\sqrt{3\tau^2-r^2} \\[4pt] \hspace{5.6cm} -r\arcsin \frac{r}{\sqrt{3}\tau}+(1+r)\arcsin \frac{r+1}{\sqrt{3}\tau}\,,
\end{matrix}
 & \frac{2(r+1)}{3} \leq \tau < \frac{2r}{\sqrt{3}} \text{ if } r>\frac{1}{\sqrt{3}-1}\,, \\[18pt]
-\frac32\tau+\sqrt{3}\tau-\sqrt{3\tau^2-(1+r)^2}+(r+1)\arccos \frac{1+r}{\sqrt{3}\tau} & \frac{2r}{\sqrt{3}}\leq \tau <\frac{2(r+1)}{3} \text{ if } r\leq\frac{1}{\sqrt{3}-1}\,, \\[10pt]
-\frac{(1+r)\pi}{6}+\sqrt{3\tau^2-(1+r)^2}-\frac32\tau +(1+r)\arcsin \frac{r+1}{\sqrt{3}\tau}\,, & \max\Big(\frac{2(r+1)}{3},\frac{2r}{\sqrt{3}}\Big) \leq \tau < \frac{2(r+1)}{\sqrt{3}} \,, \\[10pt]
\end{matrix*} \right.}
\end{equation} 
where all regimens are continuous with their neighboring ones. $I_L^{(+)}$ is derived similarly from $\mathcal{D}^{(+)}_L$, and reads
\begin{equation}
\scalemath{0.89}{
I_L^{(+)}(\tau,r)= \left \{\begin{matrix*}[l]
0 \,, & \tau < \frac{r}{\sqrt{3}} \,, \\[10pt]

\sqrt{3\tau^2-r^2} - r \arccos \frac{r}{\sqrt{3}\tau} \,, & 
\frac{r}{\sqrt{3}} \leq \tau < \min \Big(\frac{2r}{3},\frac{1+r}{\sqrt{3}}\Big) \,, \\[10pt]

\hspace{-2.5cm}
\begin{matrix}
\frac{\pi}{2} + \sqrt{3\tau^2-r^2}-\sqrt{3\tau^2-(r+1)^2} 
\\[4pt] \hspace{5cm} + r \arcsin \frac{r}{\sqrt{3}\tau} - (1+r) \arcsin \frac{1+r}{\sqrt{3}\tau} \,,
\end{matrix} & 
\frac{1+r}{\sqrt{3}} \leq \tau < \frac{2r}{3} \text{ if } r>3+2\sqrt{3} \,, \\[18pt]

-\frac{r \pi}{3} + \sqrt{3}\tau-\sqrt{3\tau^2-r^2} 
+ r \arccos \frac{r}{\sqrt{3}\tau} \,, & 
\frac{2r}{3} \leq \tau < \frac{1+r}{\sqrt{3}} \text{ else} \,, \\[10pt]

\hspace{-1cm}
\begin{matrix}
\frac{(3+r)\pi}{6} +\sqrt{3}\tau-\sqrt{3\tau^2-r^2} - \sqrt{3\tau^2-(r+1)^2} \\[4pt] \hspace{3.5cm}
- (1+r)\arcsin \frac{1+r}{\sqrt{3}\tau} + r \arccos \frac{r}{\sqrt{3}\tau} \,, 
\end{matrix}
&
\max\Big(\frac{2r}{3}, \frac{1+r}{\sqrt{3}}\Big) \leq \tau < \frac{2}{3}\sqrt{1+r+r^2} \,, \\[18pt]

-\frac{\pi}{3} + \sqrt{3}\tau-\sqrt{3\tau^2-1}
+ \arccos \frac{1}{\sqrt{3}\tau} \,, &
\frac{2}{3}\sqrt{1+r+r^2} \leq \tau < \frac{2r}{\sqrt{3}} \,, \\[10pt]

\hspace{-1.1cm}
\begin{matrix}
\frac{(1+r)\pi}{6}+\frac32\tau+\sqrt{3}\tau-\sqrt{3\tau^2-1}-\sqrt{3\tau^2-r^2} \\[4pt] \hspace{4.75cm}- \arcsin \frac{1}{\sqrt{3}\tau} 
-  r \arcsin \frac{r}{\sqrt{3}\tau}\,,
\end{matrix}
 &
\frac{2r}{\sqrt{3}} \leq \tau < \frac{2(1+r)}{\sqrt{3}} \,, \\[18pt]
\hspace{-0.4cm}
\begin{matrix}
\sqrt{3}\tau-\sqrt{3\tau^2-1} - \sqrt{3\tau^2-r^2} + \sqrt{3\tau^2-(1+r)^2} \\[4pt] \hspace{1cm}
- \arcsin \frac{1}{\sqrt{3}\tau} - r \arcsin \frac{r}{\sqrt{3}\tau} + (1+r) \arcsin \frac{1+r}{\sqrt{3}\tau} \,,
\end{matrix}
 &
\tau \geq \frac{2(1+r)}{\sqrt{3}} \,.
\end{matrix*}\right.}
\end{equation}
The results found for $I_L^{(\pm)}$ remain identical for the regions of $k$ not covered. $I_L$ thus finally reads 
\begin{equation}
I_L(\tau,r)=2\Big(I_L^{(-)}(\tau,r) +I_L^{(+)}(\tau,r)\Big).
\end{equation}

We now perform a similar derivation for $I_R$. Focusing on $k\in [\frac{\pi}{2},\frac{2\pi}{3}]$, we now invert the relations found in App.\,\ref{App:Pheno2},
\begin{equation}
\begin{aligned}
t_{e}=\frac{\ell_{B}}{\sqrt{3}\sin(k-\pi/6)} &\iff k_{e}=\frac{\pi}{6}+ \arcsin\bigg(\frac{1}{\sqrt{3}\tau}\bigg) \,,\\
t_R=\frac{-\ell_{B}}{\sqrt{3}\cos(k)} &\iff k_R=\arccos\bigg(\frac{-1}{\sqrt{3}\tau}\bigg)\,, \\
t_{L}^1=\frac{\ell_{B}}{\sqrt{3}\sin(k+\pi/6)} &\iff k_{L}^1=\frac{5\pi}{6}- \arcsin\bigg(\frac{1}{\sqrt{3}\tau} \bigg)\,, \\
t_{L}^2=\frac{-\ell_{A}}{\sqrt{3}\cos(k)} &\iff k_{L}^2=\arccos\bigg(\frac{-r}{\sqrt{3}\tau}\bigg)\,, \\
t_{F}=-\frac{\ell_{A}+\ell_{B}}{\sqrt{3}\cos(k)} &\iff k_{F}=\arccos\bigg(\frac{-(1+r)}{\sqrt{3}\tau}\bigg)\, . \\
\end{aligned}
\end{equation}
Integrating individual contributions of \eqref{piecewise2} yields
\begin{equation}
I^{(-)}_R(\tau,r)= \left \{\begin{matrix*}[l]
0 \,, & \tau < \frac{1}{\sqrt{3}}  \,, \\[10pt]
\sqrt{3\tau^2-1}-\arccos \frac{1}{\sqrt{3}\tau} \,, & \frac{1}{\sqrt{3}} \leq \tau < \frac{2}{3}\,, \\[10pt]
-\frac\pi3 +\sqrt{3}\tau-\sqrt{3\tau^2-1}+ \arccos \frac{1}{\sqrt{3}\tau}\,, & \frac{2}{3} \leq \tau < \frac{2r}{\sqrt{3}}\,, \\[10pt]
\hspace{-1.36cm}
\begin{matrix}
\frac{(r+1)\pi}{6}+\frac32\tau +\sqrt{3}\tau-\sqrt{3\tau^2-1}-\sqrt{3\tau^2-r^2}\\[4pt] \hspace{5.2cm}
-\arcsin \frac{1}{\sqrt{3}\tau}- r\arcsin \frac{r}{\sqrt{3}\tau} \,,
\end{matrix}
 & \frac{2r}{\sqrt{3}} \leq \tau < \frac{2(r+1)}{\sqrt{3}} \,, \\[18pt]

\hspace{-1.2cm}\begin{matrix}
\hspace{0.5cm}\sqrt{3}\tau-\sqrt{3\tau^2-1}-\sqrt{3\tau^2-r^2}+\sqrt{3\tau^2-(r+1)^2}\\[4pt]
\hspace{2cm}-\arcsin \frac{1}{\sqrt{3}\tau}-r\arcsin \frac{r}{\sqrt{3}\tau}+(1+r)\arcsin \frac{1+r}{\sqrt{3}\tau} \,,
\end{matrix}
 & \frac{2(r+1)}{\sqrt{3}} \leq \tau \,. \\
\end{matrix*} \right.
\end{equation}
and
\begin{equation}
\scalemath{0.90}{
I_R^{(+)}(\tau,r)= \left \{\begin{matrix*}[l]
0 \,, & \tau \leq \frac{1}{\sqrt{3}} \text{ or } \tau > \frac{2(1+r)}{\sqrt{3}} \,, \\[10pt]

\sqrt{3\tau^2-1} - \arccos \frac{1}{\sqrt{3}\tau} \,, & 
\frac{1}{\sqrt{3}} \leq \tau < \frac{2}{3} \,, \\[10pt]

 - \frac{\pi}{3}+\sqrt{3}\,\tau - \sqrt{3\tau^2-1} + \arccos \frac{1}{\sqrt{3}\tau} \,, & 
\frac{2}{3} \leq \tau < \frac{2}{3}\sqrt{1+r+r^2} \,, \\[10pt]

\hspace{-1.5cm}
\begin{matrix}
-\frac{\pi r}{3} + \sqrt{3}\tau-\sqrt{3\tau^2-r^2} - \sqrt{3\tau^2-(1+r)^2} \\[4pt]
\hspace{4.3cm}+ r \arccos \frac{r}{\sqrt{3}\tau} + (1+r) \arccos \frac{1+r}{\sqrt{3}\tau} \,,
\end{matrix}
 &
\frac{2}{3}\sqrt{1+r+r^2} \leq\tau < \min \Big(\frac{2(1+r)}{3},\frac{2r}{\sqrt{3}}\Big)  \,, \\[18pt]

-\frac32\tau + \sqrt{3}\tau - \sqrt{3\tau^2-(1+r)^2}
+ (1+r) \arccos \frac{1+r}{\sqrt{3}\tau} \,, &
\frac{2r}{\sqrt{3}}\leq \tau <\frac{2(1+r)}{3} \text{ if } r \leq \frac{1}{\sqrt{3}-1} \,, \\[10pt]
\hspace{-2.7cm}
\begin{matrix}
-\frac{\pi}{6} - \sqrt{3\tau^2-r^2} + \sqrt{3\tau^2-(1+r)^2} \\[4pt]
\hspace{5.6cm} - r \arcsin \frac{r}{\sqrt{3}\tau} + (1+r) \arcsin \frac{1+r}{\sqrt{3}\tau} \,,
\end{matrix}
 &
\frac{2(1+r)}{3} \leq\tau < \frac{2r}{\sqrt{3}} \text{ else} \,, \\[18pt]

-\frac{\pi (1+r)}{6} - \frac{3\tau}{2} + \sqrt{3\tau^2-(1+r)^2}
+ (1+r) \arcsin \frac{1+r}{\sqrt{3}\tau} \,, &
\max \Big(\frac{2(1+r)}{3},\frac{2r}{\sqrt{3}}\Big) \leq\tau < \frac{2(1+r)}{\sqrt{3}} \,.
\end{matrix*} \right.}
\end{equation}
Finally, as before $I_R(\tau,r)$ can be written as
\begin{equation}
I_R(\tau,r)=2\Big(I_R^{(-)}(\tau,r)+I_R^{(+)}(\tau,r)\Big) .
\end{equation}

\section{Periodic and open boundary conditions}

\label{App:FiniteMG}

We now turn to the study of tripartite entanglement in finite-sized systems. New mechanisms arise when considering finite-sized systems, see, e.g., \cite{Rieger:2011ipr,Blass:2012iun,Modak:2020faf,Ghasemi:2021jiy,Lagnese:2022ytv,Chalas:2024yts}. For example, entanglement undergoes periodic increases after its initial dip, called revivals, caused by the periodic course of the quasiparticles spreading in the system.

We extend here the methods used on infinite-sized systems to capture the Markov gap in the presence of revivals, both for periodic and open boundary conditions.

\subsection{Entanglement entropy}
\label{App:EEFinite}

We begin by quickly reviewing bipartite entanglement dynamics in the presence of revivals. In a periodic system of size $L$ containing a subregion $A$ of size $\ell$, consider a generalization of the scaling limit to $t,\ell,L \rightarrow \infty$, with both $t/L$ and $\ell/L$ fixed. Consequently, $A$ becomes a macroscopic region of an infinite periodic system. This isolates the quasiparticle contribution from the subleading ones while retaining the periodicity of the system. The entanglement entropy time-evolution for periodic systems reads \cite{Chapman:2018hou,Modak:2020faf}
\begin{equation}
\scalemath{0.9}{
S_\ell(t)=\int_{\big \{ \frac{2v(k)t}{L} \big \}< \frac{\ell}{L}} \frac{dk}{2\pi} s(k) L \,\bigg \{\frac{2v(k)t}{L}\bigg \} + \ell \int_{\frac{\ell}{L}\leq \big \{ \frac{2v(k)t}{L}\big\} <1-\frac{\ell}{L}} \frac{dk}{2\pi} s(k) + \int_{1-\frac{\ell}{L}\leq \big \{\frac{2v(k)t}{L}\big \}} \frac{dk}{2\pi} s(k) L \,\bigg (1-\bigg \{ \frac{2v(k)t}{L} \bigg\} \bigg) \,,}
\label{ThEqRev}
\end{equation}
where $\{x\}$ denotes the fractional part of $x$, $s(k)$ the entanglement entropy density and $v(k)$ the velocity of both quasiparticles of a pair of momentum $k$. Let us rederive this result using a more systematic formalism. 

\medskip
Introduce left- and right-movers as piecewise functions
\bea
&x_\rightarrow(k,v,t)=\left \{ \begin{matrix*}[l]
x+v(k)t \,, & 0\leq t \leq \frac{L-x}{v(k)} \,, \\
x-L+v(k)t \,, & \frac{L-x}{v(k)} < t \leq \frac{L}{v(k)} \,,
\end{matrix*} \right .
\label{xR}\\[0.5ex]
&x_\leftarrow(k,v,t)=\left \{ \begin{matrix*}[l]
x-v(k)t \,, & 0\leq t \leq \frac{x}{v(k)} \,, \\
x+L-v(k)t \,, & \frac{x}{v(k)} < t \leq \frac{L}{v(k)} \,.
\end{matrix*} \right .
\label{xL}
\eea
where $t$ is now defined mod $T_k=L/v(k)$ for the remainder of this section unless stated otherwise. Making these movers depend explicitly on $x$ rather than $t$, with the additional constraint $x\in[0,L]$, is easier to handle, yielding
\begin{equation}
x_\rightarrow(k,t)=\left \{ \begin{matrix*}[l]
x+v(k)t \,, & 0\leq x \leq L-v(k)t \,, \\
x-L+v(k)t \,, & L-v(k)t < x \leq L \,,
\end{matrix*} \right .
\label{x_R}
\end{equation}

\begin{equation}
x_\leftarrow(k,t)=\left \{ \begin{matrix*}[l]
x+L-v(k)t \,, & 0 \leq x \leq v(k)t \,, \\
x-v(k)t \,, &  v(k)t < x \leq L \,,
\end{matrix*} \right . 
\label{x_L}
\end{equation}
where we omit the dependency on $v$ since for a given $k$, both quasiparticles are equal here. We choose the subregion $A$ to be the segment $[0,\ell]$, and assume without loss of generality that $\ell\leq L/2$. Finally, we assign the left-moving quasiparticle to $A$ and the right-moving one to $C$, and derive the set of conditions for non-zero entanglement from quasiparticles originating from a point $x$ at $t=0$ as
\bea
&x_{\rightarrow}(k,t)\leq L \,, \quad x_\leftarrow(k,t)\leq \ell \,, \quad (\alpha) \\[1ex]
&\,\spp x_\rightarrow(k,t)\geq \ell \,, \quad x_\leftarrow(k,t) \geq 0 \,. \quad \spp(\beta)
\eea
We set to integrate the indicator function that signals $x\in(\alpha) \cap(\beta)$ for $x\in [0,L]$. Since $x_{\rightarrow}$ and $x_\leftarrow$ both have two regimes, we have that
\begin{equation}
    x\in [0,L]\cap (\alpha) \cap(\beta) \iff x\in[\mathcal{L},\mathcal{R}]
\end{equation}
with 
\bea
&\mathcal{R}=\left \{ \begin{matrix*}[l]
\min\big(L-v(k)t,\ell+v(k)t-L\big) \,, & 0\leq x \leq \min\big(v(k)t,L-v(k)t\big) \,, \\[2pt]
\min\big(L-v(k)t,\ell+v(k)t\big) \,, & v(k)t\leq x \leq L-v(k)t \,, \\[2pt]
\min\big(2L-v(k)t,\ell+v(k)t-L\big) \,, & L-v(k)t \leq x \leq v(k)t \,, \\[1pt]
\min\big(2L-v(k)t,\ell+v(k)t\big) \,, & \max\big(v(k)t,L-v(k)t\big)\leq x \leq L \,, 
\end{matrix*} \right . 
\label{bipartiteMax}\\[0.5ex]
&\mathcal{L}=\left \{ \begin{matrix*}[l]
\max\big(\ell-v(k)t,v(k)t-L\big) \,, & 0\leq x \leq \min\big(v(k)t,L-v(k)t\big) \,, \\[2pt]
\max\big(\ell-v(k)t,v(k)t\big) \,, & v(k)t\leq x \leq L-v(k)t \,, \\[2pt]
\max\big(\ell+L-v(k)t,v(k)t-L\big) \,, & L-v(k)t \leq x \leq v(k)t \,, \\[2pt]
\max\big(\ell+L-v(k)t,v(k)t\big) \,, & \max \big(v(k)t,L-v(k)t\big)\leq x \leq L \,. 
\end{matrix*} \right .
\label{bipartiteMin}
\eea
Integration of the indicator functions then yields
\begin{equation}
\scalemath{0.9}{
\begin{aligned}
\int_0^L dx \, \mathbf1_{[\mathcal{L},\mathcal{R}]}(x)  &= \int_0^{v(k)t} dx\,\mathbf1_{[\mathcal{L},\mathcal{R}]}(x)  + \int_{v(k)t}^{L-v(k)t} dx \, \mathbf1_{[\mathcal{L},\mathcal{R}]}(x)  + \int_{L-v(k)t}^L dx \, \mathbf1_{[\mathcal{L},\mathcal{R}]}(x) \\
& =  0 + \max \big [ 0,\min\big( L-\ell,\ell,L-2v(k)t,2v(k)t \big) \big]+0 \\
& = \max \big [ 0,\min\big(\ell,L-2v(k)t,2v(k)t \big) \big],
\end{aligned}}
\label{PhenoXBipartite}
\end{equation}
where we used $\ell<L/2 \iff \ell < L-\ell$ in the last equality. Note that we chose $t\leq T_k/2$ here such that $v(k)t\leq L-v(k)t$. Taking $t > T_k/2$ is equivalent since the quasiparticle pair has periodicity $T_k/2$.

Finally, integrating \eqref{PhenoXBipartite} with respect to $k\in [0,\pi]$ and enforcing explicitely the $T_k/2$-time periodicity one obtains 
\begin{equation}
\begin{aligned}
S_\ell(t)= 2\int_{0}^\pi \frac{dk}{2\pi}s(k) \min \bigg [\ell,L\bigg \{\frac{2v(k)t}{L}\bigg \},L \bigg(1-\bigg\{\frac{2v(k)t}{L}\bigg\}\bigg) \bigg ] \,,
\end{aligned}
\end{equation}
where the factor of two accounts for the other choice of quasiparticles assignment (left-moving one to $C$ and right-moving one to $A$). Here, $t$ is no longer defined $\text{mod }  T_k$; instead, periodicity is enforced by the fractional part functions. This formula is precisely equal to \eqref{ThEqRev}.

\subsection{Tripartite entanglement: periodic systems}

The tripartite case can be derived analogously with the formalism introduced in the previous section. We do not choose a particular entanglement density, so the calculations apply to any tripartite entanglement measure. We begin by considering a region $A-B-C$ with $A=[0,\ell_A)$, $B=[\ell_A,\ell_A+\ell_B)$ and $C=[\ell_A+\ell_B,L)$, where $L$ is the size of the entire system. We first work in the region $k\in [\pi/3,2\pi/3]$. Since each quasiparticle has a different velocity, we introduce $t_i(k)=t \mod L/v_i(k) \,,\; i=R,M,L$, the periodic time of each quasiparticles. They will be defined as $x_R(k,t)=x_\rightarrow(k,v_R,t_R)$, $x_M(k,t)=x_\leftarrow(k,|v_M|,t_M)$ and $x_L(k,t)=x_\leftarrow(k,|v_L|,t_L)$. Consider a single source of quasiparticles located at $x$. The set of conditions for non-zero \mbox{entanglement reads}
\bea
&x_R(k,t)\leq L\,, \quad x_M(k,t) \leq \ell_A+\ell_B\,, \quad x_L(k,t) \leq \ell_A \,,
\label{Cond1}\\[.5em]
&x_R(k,t)\geq \ell_A+\ell_B \,, \quad x_M(k,t) \geq \ell_A \,, \quad x_L(k,t) \geq 0 \,.
\label{Cond2}
\eea
The regions $\mathcal{L}$ and $\mathcal{R}$ follow as
\bea\label{xmax}
&\scalemath{0.85}{
\mathcal{R}=\left \{ \begin{matrix*}[l]
\min\big(L-v_R(k)t_R,\ell_A+\ell_B-L+v_M(k)t_M,\ell_A-L+v_L(k)t_L\big) \,, & 0\leq x \leq \min\big(L-v_R(k)t_R,v_M(k)t_M,v_L(k)t_L\big) \,, \\[2pt]
\min\big(L-v_R(k)t_R,\ell_A+\ell_B-L+v_M(k)t_M,\ell_A+v_L(k)t_L\big) \,, & v_L(k)t_L\leq x \leq \min\big(L-v_R(k)t_R,v_M(k)t_M\big) \,, \\[2pt]
\min\big(L-v_R(k)t_R,\ell_A+\ell_B+v_M(k)t_M,\ell_A-L+v_L(k)t_L\big) \,, & v_M(k)t_M\leq x \leq \min\big(L-v_R(k)t_R,v_L(k)t_L\big) \,, \\[2pt]
\min\big(L-v_R(k)t_R,\ell_A+\ell_B+v_M(k)t_M,\ell_A+v_L(k)t_L\big) \,, & \max\big(v_M(k)t_M,v_L(k)t_L\big)\leq x \leq L-v_R(k)t_R \,, \\[2pt]
\min\big(2L-v_R(k)t_R,\ell_A+\ell_B-L+v_M(k)t_M,\ell_A-L+v_L(k)t_L\big) \,, & L-v_R(k)t_R\leq x \leq \min\big(v_M(k)t_M,v_L(k)t_L\big) \,, \\[2pt]
\min\big(2L-v_R(k)t_R,\ell_A+\ell_B-L+v_M(k)t_M,\ell_A+v_L(k)t_L\big) \,, & \max\big(L-v_R(k)t_R,v_L(k)t_L\big)\leq x \leq v_M(k)t_M \,, \\[2pt]
\min\big(2L-v_R(k)t_R,\ell_A+\ell_B+v_M(k)t_M,\ell_A-L+v_L(k)t_L\big) \,, & \max\big(L-v_R(k)t_R,v_M(k)t_M\big)\leq x \leq v_L(k)t_L \,, \\[2pt]
\min\big(2L-v_R(k)t_R,\ell_A+\ell_B+v_M(k)t_M,\ell_A+v_L(k)t_L\big) \,, & \max\big(L-v_R(k)t_R,v_M(k)t_M,v_L(k)t_L\big)\leq x \leq L \,, \\[2pt]
\end{matrix*} \right . }\\
&\scalemath{0.85}{
\mathcal{L}=\left \{ \begin{matrix*}[l]
\max\big(\ell_A+\ell_B-v_R(k)t_R,\ell_A-L+v_M(k)t_M,-L+v_L(k)t_L\big) \,, & 0\leq x \leq \min\big(L-v_R(k)t_R,v_M(k)t_M,v_L(k)t_L\big) \,, \\[2pt]
\max\big(\ell_A+\ell_B-v_R(k)t_R,\ell_A-L+v_M(k)t_M,v_L(k)t_L\big) \,, & v_L(k)t_L\leq x \leq \min\big(L-v_R(k)t_R,v_M(k)t_M\big) \,, \\[2pt]
\max\big(\ell_A+\ell_B-v_R(k)t_R,\ell_A+v_M(k)t_M,-L+v_L(k)t_L\big) \,, & v_M(k)t_M\leq x \leq \min\big(L-v_R(k)t_R,v_L(k)t_L\big) \,, \\[2pt]
\max\big(\ell_A+\ell_B-v_R(k)t_R,\ell_A+v_M(k)t_M,v_L(k)t_L\big) \,, & \max\big(v_M(k)t_M,v_L(k)t_L\big)\leq x \leq L-v_R(k)t_R \,, \\[2pt]
\max\big(\ell_A+\ell_B+L-v_R(k)t_R,\ell_A-L+v_M(k)t_M,-L+v_L(k)t_L\big) \,, & L-v_R(k)t_R\leq x \leq \min\big(v_M(k)t_M,v_L(k)t_L\big) \,, \\[2pt]
\max\big(\ell_A+\ell_B+L-v_R(k)t_R,\ell_A-L+v_M(k)t_M,v_L(k)t_L\big) \,, & \max\big(L-v_R(k)t_R,v_L(k)t_L\big)\leq x \leq v_M(k)t_M \,, \\[2pt]
\max\big(\ell_A+\ell_B+L-v_R(k)t_R,\ell_A+v_M(k)t_M,-L+v_L(k)t_L\big) \,, & \max\big(L-v_R(k)t_R,v_M(k)t_M\big)\leq x \leq v_L(k)t_L \,, \\[2pt]
\max\big(\ell_A+\ell_B+L-v_R(k)t_R,\ell_A+v_M(k)t_M,v_L(k)t_L\big) \,, & \max\big(L-v_R(k)t_R,v_M(k)t_M,v_L(k)t_L\big)\leq x \leq L \,. \\[2pt]
\end{matrix*} \right . }
\label{xmin}
\eea
As was seen in the bipartite case, these functions define several mutually exclusive regimes depending on the ordering between $L-v_R(k)t_R,v_M(k)t_M$, and $v_L(k)t_L$. One then integrates the indicator functions of $x\in [\mathcal{L},\mathcal{R}]$ continuously for each ordering. The final expression reads
\begin{equation}
\scalemath{0.95}{
\begin{aligned}
S^{(L,M,R)}_{\ell_A,\ell_B,\ell_C}(t)&=\int_{\begin{matrix} \\[0.35pt]
1-\big \{\frac{v_Rt}{L}\big \}\leq \big \{\frac{v_Mt}{L} \big\} \leq \big \{\frac{v_Lt}{L} \big \}
\end{matrix}}^{k\in [\pi/3,2\pi/3]} \hspace{-1.5cm} (dk/2\pi) \, s(k)\spp \max\spm\Big[0,\min\big(L-v_Rt_R,\ell_A+\ell_B-L+v_Mt_M,\ell_A-L+v_Lt_L\big) \\[-10.pt]& \hspace{6.5cm}-\max\big(\ell_A+\ell_B-v_Rt_R,\ell_A-L+v_Mt_M,-L+v_Lt_L\big) \Big] \\
& +  \int_{\begin{matrix} \\[0.35pt]
\big \{\frac{v_Mt}{L}\big \}\leq \big \{\frac{v_Lt}{L} \big\} \leq 1-\big \{\frac{v_Rt}{L} \big \}
\end{matrix}}^{k\in [\pi/3,2\pi/3]} \hspace{-1.5cm} (dk/2\pi) \, s(k)\spp\max\spm\Big[0,\min\big(L-v_Rt_R,\ell_A+\ell_B+v_Mt_M,\ell_A+v_Lt_L\big) \\[-10.pt]& \hspace{6.5cm}-\max\big(\ell_A+\ell_B-v_Rt_R,\ell_A+v_Mt_M,v_L(k)t_L\big)\Big] \\
& + \int_{\begin{matrix} \\[0.35pt]
\big \{\frac{v_Lt}{L}\big \}\leq 1-\big \{\frac{v_Rt}{L} \big\} \leq \big \{\frac{v_Mt}{L} \big \}
\end{matrix}}^{k\in [\pi/3,2\pi/3]} \hspace{-1.5cm} (dk/2\pi) \, s(k)\spp\max\spm\Big[0,\min\big(L-v_R(k)t_R,\ell_A+\ell_B-L+v_Mt_M,\ell_A+v_Lt_L\big) \\[-10.pt] &\hspace{6.5cm} -\max\big(\ell_A+\ell_B-v_Rt_R,\ell_A-L+v_Mt_M,v_Lt_L\big)\Big] \,,
\end{aligned}}
\label{S123rev}
\end{equation}
where $(L,M,R)$ stand for the respective assignment of each quasiparticle to the subregions $(A,B,C)$. Other configurations are obtained by rotation of the periodic system. Notice first that for 3-cycles of $(L,M,R)$, the change of assignment is equivalent to a rotation of the periodic system. This in turn amounts to a 3-cycle of region lengths in \eqref{S123rev}. Permutations of $(L,M,R)$ and any equivalent 3-cycles can be recovered by taking the mirror image of the original system, whereby we make the change $\big(L-v_R(k)t_R,v_M(k)t_M,v_L(k)t_L\big) \leftrightarrow \big( v_R(k)t_R,L-v_M(k)t_M,L-v_L(k)t_L \big)$ in \eqref{S123rev}. In other words, we changed every right-mover into a left-mover and vice versa. We label these mirror images $\tilde{S}^{(L,M,R)}_{\ell_A,\ell_B,\ell_C}$. Finally, one can access the region $k\in [2\pi/3,\pi]$ by mirroring the system without mirroring the quasiparticles' propagation (i.e. permuting all the first and last subregions). The final expression then reads
\begin{equation}
S_{\ell_A,\ell_B,\ell_C}(t)=\sum_{(\alpha,\beta,\gamma) \in S_3} \Big(S_{\ell_\alpha,\ell_\beta,\ell_\gamma}^{(L,M,R)}(t) + \tilde{S}_{\ell_\alpha,\ell_\beta,\ell_\gamma}^{(L,M,R)}(t) \Big),
\label{LentTriRev}
\end{equation}
where $S_3$ is the symmetric group on three elements $(A,B,C)$. Note that at this point, this is still a general tripartite entanglement measure, as we did not specify any density $s(k)$. In the following, we consider $s(k)=m_{A:B}^{(3,1)}$.

\begin{figure}[b]
\centering
\hspace{-0.5cm}
\includegraphics[scale=1]{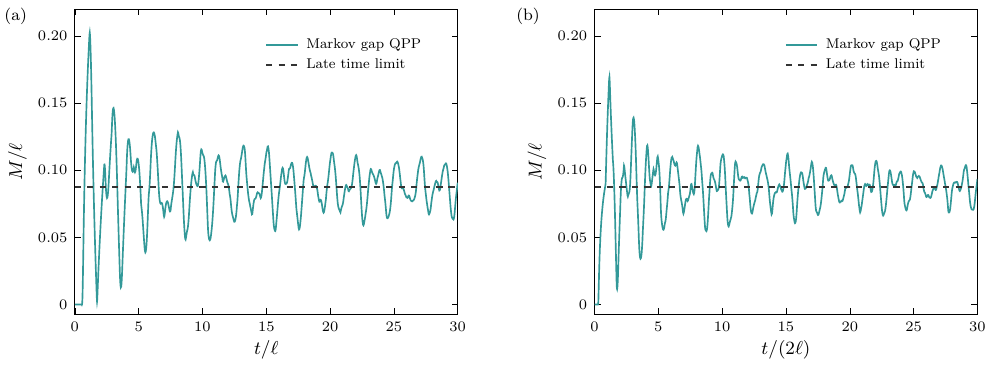}
\vspace{-5pt}
\caption{Markov gap evolution after a $|C_{3,1}\ra$ quench for
$\ell_A=\ell_B=\ell_C=\ell$ with (a) periodic and (b) open boundary conditions.}
\label{Fig:EntEvFinite}
\end{figure}

Using \eqref{LentTriRev}, we present in Fig.\,\ref{Fig:EntEvFinite} the Markov gap evolution after a $|C_{3,1}\ra$ quench for finite systems with periodic and open boundary conditions. We observe the familiar initial growth of entanglement found in infinite systems, followed by oscillations from winding or reflected trajectories. At a fixed generic momentum the relative phases are quasiperiodic. Assuming that at a large enough $t$, quasiparticles are distributed randomly along the system, one finds 
\begin{equation}
M_\infty^{}(\ell_A,\ell_B,\ell_C)=6m_{A:B}^{(3,1)}\frac{\ell_A \ell_B \ell_C}{L^3}\frac{L2\pi/3}{2\pi}=2m_{A:B}^{{(3,1)}}\frac{\ell_A\ell_B\ell_C}{L^2} \,,
\label{SInfRev}
\end{equation}
where $\ell_A\ell_B\ell_C/L^3$ represents the probabilities of each QP to be located in its assigned region when distributed at random, six is the number of configurations of assignments of $(L,M,R)$ to any permutation of $(A,B,C)$, and $L(2\pi/3)/2\pi$ is the number of quasiparticle triplets in the system. Our results are in perfect agreement with numerical simulations in the hydrodynamic regime.

\subsection{Open systems}

\begin{figure}[b]
\centering
\includegraphics[scale=1.3]{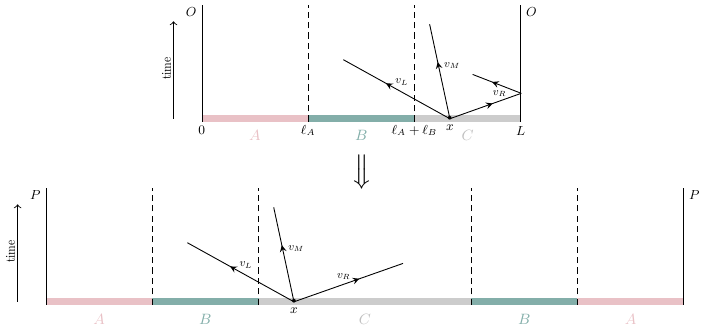}
\caption{Tripartite Entanglement evolution in the quasiparticles picture for two adjacent regions $A$ and $B$ in a finite system with open boundary conditions (denoted $O$ at each boundary). We unfold the open boundary region to make a region of size $2L$ with periodic boundary conditions (denoted $P$ at each boundary). The periodic region can then be shifted at will to compute the entanglement evolution between every subsystems.}
\label{Fig:TriEntOBC}
\end{figure}

For open boundary conditions, quasiparticles bounce off of domain walls \cite{Rieger:2011ipr}. In the tripartite case, we make use of a folding trick to simplify our calculations, see Fig.\,\ref{Fig:TriEntOBC}. Indeed, taking a mirrored copy of the original system and placing it next to it creates a $2L$ system with periodic boundary conditions. However, this creates two subregions $A$ and $B$ (the subregions of $C$ are directly next to one another after this operation). One would then have to compute the entanglement evolution for each assignment of  $R,M,L\in \{A_1,A_2,B_1,B_2,C\}$. This makes the computations very cumbersome, although entirely possible. We avoid this by shifting the entire region by $\ell_A$ to bundle both $A$ regions. Focusing first on the leftmost $B$ region, we introduce $a_i(k,t)=L\big[1+\text{sign}\big(v_i(k)\big)\big] -v_i(k)t, \; i=L,M,R$ and compute the entanglement evolution as previously. This yields
\begin{equation}
\scalemath{0.95}{
\begin{aligned}
S_{\ell_A,\ell_B,\ell_C}^{(i,j,l)}(t)&=\int_{\begin{matrix} \\[0.5pt]
a_i(k,t) \leq a_j(k,t) \leq a_l(k,t)
\end{matrix}}^{k\in [\pi/3,\pi]} \hspace{-3.5cm} (dk/2\pi) \, s(k) \max \big [0,\min\big(-\ell_B+a_i(k,t),2\ell_A+\ell_B-2L+a_j(k,t),2\ell_A-2L+a_l(k,t)\big) \\& \hspace{2cm}-\max\big(2\ell_A+\ell_B-2L+a_i(k,t),2\ell_A-2L+a_j(k,t),-2L+a_l(k,t)\big) \big ] \\
& +  \int_{\begin{matrix} \\[0.5pt]
a_j(k,t) \leq a_l(k,t) \leq a_i(k,t)
\end{matrix}}^{k\in [\pi/3,\pi]} \hspace{-3.5cm} (dk/2\pi) \, s(k) \max \big [0,\min\big(-\ell_B+a_i(k,t),2\ell_A+\ell_B+a_j(k,t),2\ell_A+a_l(k,t)\big) \\& \hspace{2.5cm}-\max\big(2\ell_A+\ell_B-2L+a_i(k,t),2\ell_A+a_j(k,t),a_l(k,t)\big)\big] \\
& + \int_{\begin{matrix} \\[0.5pt]
a_l(k,t) \leq a_i(k,t) \leq a_j(k,t)
\end{matrix}}^{k\in [\pi/3,\pi]} \hspace{-3.5cm} (dk/2\pi) \, s(k) \max\big[0,\min\big(-\ell_B+a_i(k,t),2\ell_A+\ell_B-2L+a_j(k,t),2\ell_A+a_l(k,t)\big) \\ &\hspace{3cm} -\max\big(2\ell_A+\ell_B-2L+a_i(k,t),2\ell_A-2L+a_j(k,t),a_l(k,t)\big)\big]\,. \\
\end{aligned}}
\label{S123OBC}
\end{equation}
The contribution from the rightmost $B$ regions is obtained by shifting the system a second time, which amounts to exchanging the regions $A$ and $C$. Exchanging $\ell_A$ and $\ell_C$ in the previous expression thus yields the desired result. 

Finally, we sum over all assignments $(L,M,R) \in (A,B,C)$ and find
\begin{equation}
S_{\ell_A,\ell_B,\ell_C}(t)=\frac{1}{2}\sum_{(i,j,l)\in S_3} \Big( S_{\ell_A,\ell_B,\ell_C}^{(i,j,l)}(t)+S_{\ell_C,\ell_B,\ell_A}^{(i,j,l)}(t) \Big) \,.
\end{equation}
where $S_3$ is the symmetric group of degree of $(L,M,R)$, and we have taken half of the final expression to account for the doubled system. An example of the Markov gap evolution in an open system can be found in Fig.\,\ref{Fig:EntEvFinite}. It exhibits similar features to the periodic case.

\section{Time-dependent correlation functions used for numerical calculations}
\lb{Apdx:correl}

Since all the states we consider are Gaussian, the time evolution after the quench is completely characterized by the two-point correlation matrix, which in position-space is
\bea\lb{correl_def}
G_{ij}(t) = \la\psi_0|e^{itH} c_i^\dagger c_j e^{-itH}|\psi_0\ra\spp.
\eea
The fermion operator in the Heisenberg picture reads
\bea
e^{itH} c_j^\dagger e^{-itH} = \frac1L \sum_{k,l}e^{-i(j-l)k}e^{it\eps_k} c_l^\dagger\spp.
\eea
Plugging this into \eqref{correl_def} yields
\bea
\lb{correl_gen}
G_{ij}(t) = \frac{1}{L^2} \sum_{\substack{k,k' \\ l,l'}}e^{-i(i-l)k}e^{i(j-l')k'}e^{it(\eps_k-\eps_{k'})}G_{ll'}(0)\spp.
\eea

In the thermodynamic limit $L\rightarrow\infty$, the time-dependent correlation matrix for an initial crystal state reads \cite{Berthiere:2024sio}
\bea
&G_{ij}(t) = G_{ij}(\infty) + \frac{i^{j-i}}p\spm\spm \sum_{n\scalebox{0.7}{$\in$} \omega_{p,q}}\spm\spm a_{i,j}^n\spp J_{j-i}\big(2t\sin(\pi n/p)\big)\spp,
\eea
where $J_\alpha(z)$ is Bessel's function of the first kind, and 
\bea
&G_{ij}(\infty) = \frac qp\delta_{i,j}\spp,\nn\\
&a_{i,j}^n 
 =\frac{\sin\spm\big(\spm\frac{n\pi q\spm}{p}\big)}{\sin\spm\big(\spm\frac{n\pi}{p}\spm\big)}\cos\spm\Big(\frac\pi2(i-j) + \frac{\pi n}{p} (i + j+q -1)\spm\Big),\\
&\omega_{p,q} =[1,p-1]\backslash \Big\{d,2d,\cdots,p-d\Big\},\nn
\eea
and we recall that $g=\gcd(p,q)$ and $d=p/g$. 

For instance, the correlation matrix corresponding to a quench from $|C_{3,1}\ra$ reads
\bea
&G_{ij}(t) = \frac 13\delta_{ij} + \frac{2(-1)^j}{3i^{i-j}}\cos\Big(\frac{\pi}{6}(i+j)\Big)\spp J_{i-j}\big(t\sqrt{3}\big)\spp.
\eea


\counterwithout{equation}{section}
\addtocounter{equation}{-1}


\let\oldaddcontentsline\addcontentsline
\renewcommand{\addcontentsline}[3]{}
\bibliography{biblio_multipartite_crystal_quenches.bib}

@ARTICLE{1970PhRvA...2.1075B,
       author = {{Barouch}, Eytan and {McCoy}, Barry M. and {Dresden}, Max},
        title = "{Statistical Mechanics of the XY Model. I}",
      journal = {Phys.\,Rev.\,A},
         year = 1970,
       volume = {2},
       number = {3},
        pages = {1075-1092},
          doi = {10.1103/PhysRevA.2.1075},
       adsurl = {https://ui.adsabs.harvard.edu/abs/1970PhRvA...2.1075B}
}

@article{Calabrese:2007rg,
    author = "Calabrese, Pasquale and Cardy, John",
    title = "{Quantum Quenches in Extended Systems}",
    eprint = "0704.1880",
    archivePrefix = "arXiv",
    doi = "10.1088/1742-5468/2007/06/P06008",
    journal = "J. Stat. Mech.",
    volume = "0706",
    pages = "P06008",
    year = "2007"
}

@article{Calabrese:2016psf,
    author = "Calabrese, Pasquale and Essler, Fabian H. L. and Mussardo, Giuseppe",
    title = "{Introduction to \textquoteleft{}Quantum Integrability in Out of Equilibrium Systems\textquoteright{}}",
    doi = "10.1088/1742-5468/2016/06/064001",
    journal = "J. Stat. Mech.",
    volume = "2016",
    number = "6",
    pages = "064001",
    year = "2016"
}

@article{Polkovnikov:2010yn,
    author = "Polkovnikov, Anatoli and Sengupta, Krishnendu and Silva, Alessandro and Vengalattore, Mukund",
    title = "{Nonequilibrium dynamics of closed interacting quantum systems}",
    eprint = "1007.5331",
    archivePrefix = "arXiv",
    doi = "10.1103/RevModPhys.83.863",
    journal = "Rev. Mod. Phys.",
    volume = "83",
    pages = "863",
    year = "2011"
}

@article{Gogolin:2015gts,
    author = "Gogolin, Christian and Eisert, Jens",
    title = "{Equilibration, thermalisation, and the emergence of statistical mechanics in closed quantum systems}",
    eprint = "1503.07538",
    archivePrefix = "arXiv",
    doi = "10.1088/0034-4885/79/5/056001",
    journal = "Rept. Prog. Phys.",
    volume = "79",
    number = "5",
    pages = "056001",
    year = "2016"
}

@article{Essler:2016ufo,
    author = "Essler, Fabian H. L. and Fagotti, Maurizio",
    title = "{Quench dynamics and relaxation in isolated integrable quantum spin chains}",
    eprint = "1603.06452",
    archivePrefix = "arXiv",
    doi = "10.1088/1742-5468/2016/06/064002",
    journal = "J. Stat. Mech.",
    volume = "1606",
    number = "6",
    pages = "064002",
    year = "2016"
}

@article{Balasubramanian:2025jhq,
    author = "Balasubramanian, Vijay and Jiang, Hanzhi and Ross, Simon F.",
    title = "{Time evolution of multi-party entanglement signals}",
    eprint = "2511.16729",
    archivePrefix = "arXiv",
    doi = "10.1007/JHEP06(2026)055",
    journal = "JHEP",
    volume = "06",
    pages = "055",
    year = "2026"
}

@article{Maric:2024hzo,
    author = "Mari{\'c}, Vanja and Ferro, Florent and Fagotti, Maurizio",
    title = "{Disorder-order interface propagating over the ferromagnetic ground state in the transverse field Ising chain}",
    eprint = "2411.04089",
    archivePrefix = "arXiv",
    doi = "10.1103/PhysRevB.111.205118",
    journal = "Phys. Rev. B",
    volume = "111",
    number = "20",
    pages = "205118",
    year = "2025"
}

@article{Carollo:2022lrq,
    author = "Carollo, Federico and Alba, Vincenzo",
    title = "{Entangled multiplets and spreading of quantum correlations in a continuously monitored tight-binding chain}",
    eprint = "2206.07806",
    archivePrefix = "arXiv",
    doi = "10.1103/PhysRevB.106.L220304",
    journal = "Phys. Rev. B",
    volume = "106",
    number = "22",
    pages = "L220304",
    year = "2022"
}

@article{Horodecki:2009zz,
    author = "Horodecki, Ryszard and Horodecki, Pawel and Horodecki, Michal and Horodecki, Karol",
    title = "{Quantum entanglement}",
    eprint = "quant-ph/0702225",
    archivePrefix = "arXiv",
    doi = "10.1103/RevModPhys.81.865",
    journal = "Rev. Mod. Phys.",
    volume = "81",
    pages = "865--942",
    year = "2009"
}

@article{Guhne:2008qic,
    author = {G{\"u}hne, Otfried and T{\'o}th, G{\'e}za},
    title = "{Entanglement detection}",
    eprint = "0811.2803",
    archivePrefix = "arXiv",
    doi = "10.1016/j.physrep.2009.02.004",
    journal = "Phys. Rept.",
    volume = "474",
    pages = "1--75",
    year = "2009"
}

@article{Harper:2024ker,
    author = "Harper, Jonathan and Takayanagi, Tadashi and Tsuda, Takashi",
    title = "{Multi-entropy at low Renyi index in 2d CFTs}",
    eprint = "2401.04236",
    archivePrefix = "arXiv",
    reportNumber = "YITP-24-02",
    doi = "10.21468/SciPostPhys.16.5.125",
    journal = "SciPost Phys.",
    volume = "16",
    number = "5",
    pages = "125",
    year = "2024"
}

@article{Yuan:2024yfg,
    author = "Yuan, Ma-Ke and Li, Mingyi and Zhou, Yang",
    title = "{Reflected Multientropy and Its Holographic Dual}",
    eprint = "2410.08546",
    archivePrefix = "arXiv",
    doi = "10.1103/76vs-rxcs",
    journal = "Phys. Rev. Lett.",
    volume = "135",
    number = "9",
    pages = "091604",
    year = "2025"
}

@article{Dutta:2019gen,
    author = "Dutta, Souvik and Faulkner, Thomas",
    title = "{A canonical purification for the entanglement wedge cross-section}",
    eprint = "1905.00577",
    archivePrefix = "arXiv",
    doi = "10.1007/JHEP03(2021)178",
    journal = "JHEP",
    volume = "03",
    pages = "178",
    year = "2021"
}

@article{Kudler-Flam:2020url,
    author = "Kudler-Flam, Jonah and Kusuki, Yuya and Ryu, Shinsei",
    title = "{Correlation measures and the entanglement wedge cross-section after quantum quenches in two-dimensional conformal field theories}",
    eprint = "2001.05501",
    archivePrefix = "arXiv",
    reportNumber = "YITP-20-05",
    doi = "10.1007/JHEP04(2020)074",
    journal = "JHEP",
    volume = "04",
    pages = "074",
    year = "2020"
}

@article{Bueno:2020vnx,
    author = "Bueno, Pablo and Casini, Horacio",
    title = "{Reflected entropy, symmetries and free fermions}",
    eprint = "2003.09546",
    archivePrefix = "arXiv",
    doi = "10.1007/JHEP05(2020)103",
    journal = "JHEP",
    volume = "05",
    pages = "103",
    year = "2020"
}

@article{Li:2020ceg,
    author = "Li, Tianyi and Chu, Jinwei and Zhou, Yang",
    title = "{Reflected Entropy for an Evaporating Black Hole}",
    eprint = "2006.10846",
    archivePrefix = "arXiv",
    doi = "10.1007/JHEP11(2020)155",
    journal = "JHEP",
    volume = "11",
    pages = "155",
    year = "2020"
}

@article{Berthiere:2020ihq,
    author = "Berthiere, Cl{\'e}ment and Chen, Hongjie and Liu, Yuefeng and Chen, Bin",
    title = "{Topological reflected entropy in Chern-Simons theories}",
    eprint = "2008.07950",
    archivePrefix = "arXiv",
    doi = "10.1103/PhysRevB.103.035149",
    journal = "Phys. Rev. B",
    volume = "103",
    number = "3",
    pages = "035149",
    year = "2021"
}

@article{Bueno:2020fle,
    author = "Bueno, Pablo and Casini, Horacio",
    title = "{Reflected entropy for free scalars}",
    eprint = "2008.11373",
    archivePrefix = "arXiv",
    doi = "10.1007/JHEP11(2020)148",
    journal = "JHEP",
    volume = "11",
    pages = "148",
    year = "2020"
}

@article{Akers:2022max,
    author = "Akers, Chris and Faulkner, Thomas and Lin, Simon and Rath, Pratik",
    title = "{The Page curve for reflected entropy}",
    eprint = "2201.11730",
    archivePrefix = "arXiv",
    doi = "10.1007/JHEP06(2022)089",
    journal = "JHEP",
    volume = "06",
    pages = "089",
    year = "2022"
}

@article{Vasli:2022kfu,
    author = "Vasli, Mohammad Javad and Mohammadi Mozaffar, M. Reza and Babaei Velni, Komeil and Sahraei, Mohammad",
    title = "{Holographic study of reflected entropy in anisotropic theories}",
    eprint = "2207.14169",
    archivePrefix = "arXiv",
    doi = "10.1103/PhysRevD.107.026012",
    journal = "Phys. Rev. D",
    volume = "107",
    number = "2",
    pages = "026012",
    year = "2023"
}

@article{Afrasiar:2022fid,
    author = "Afrasiar, Mir and Basak, Jaydeep Kumar and Chandra, Ashish and Sengupta, Gautam",
    title = "{Reflected entropy for communicating black holes. Part I. Karch-Randall braneworlds}",
    eprint = "2211.13246",
    archivePrefix = "arXiv",
    doi = "10.1007/JHEP02(2023)203",
    journal = "JHEP",
    volume = "02",
    pages = "203",
    year = "2023"
}

@article{Berthiere:2023gkx,
    author = "Berthiere, Cl{\'e}ment and Parez, Gilles",
    title = "{Reflected entropy and computable cross-norm negativity: Free theories and symmetry resolution}",
    eprint = "2307.11009",
    archivePrefix = "arXiv",
    doi = "10.1103/PhysRevD.108.054508",
    journal = "Phys. Rev. D",
    volume = "108",
    number = "5",
    pages = "054508",
    year = "2023"
}

@article{Siva:2021cgo,
    author = "Siva, Karthik and Zou, Yijian and Soejima, Tomohiro and Mong, Roger S. K. and Zaletel, Michael P.",
    title = "{Universal tripartite entanglement signature of ungappable edge states}",
    eprint = "2110.11965",
    archivePrefix = "arXiv",

    doi = "10.1103/PhysRevB.106.L041107",
    journal = "Phys. Rev. B",
    volume = "106",
    number = "4",
    pages = "L041107",
    year = "2022"
}

@article{Akers:2019gcv,
    author = "Akers, Chris and Rath, Pratik",
    title = "{Entanglement Wedge Cross Sections Require Tripartite Entanglement}",
    eprint = "1911.07852",
    archivePrefix = "arXiv",
    
    doi = "10.1007/JHEP04(2020)208",
    journal = "JHEP",
    volume = "04",
    pages = "208",
    year = "2020"
}

@article{Hayden:2021gno,
    author = "Hayden, Patrick and Parrikar, Onkar and Sorce, Jonathan",
    title = "{The Markov gap for geometric reflected entropy}",
    eprint = "2107.00009",
    archivePrefix = "arXiv",
    doi = "10.1007/JHEP10(2021)047",
    journal = "JHEP",
    volume = "10",
    pages = "047",
    year = "2021"
}

@article{Liu:2024ulq,
    author = "Liu, Bowei and Zhang, Junjia and Ohyama, Shuhei and Kusuki, Yuya and Ryu, Shinsei",
    title = "{Multiwavefunction overlap and multientropy for topological ground states in (2+1) dimensions}",
    eprint = "2410.08284",
    archivePrefix = "arXiv",

    reportNumber = "RIKEN-iTHEMS-Report-24 KYUSHU-HET-297",
    doi = "10.1103/tjcf-yryh",
    journal = "Phys. Rev. B",
    volume = "112",
    number = "12",
    pages = "125160",
    year = "2025"
}

@article{Iizuka:2025ioc,
    author = "Iizuka, Norihiro and Nishida, Mitsuhiro",
    title = "{Genuine multientropy and holography}",
    eprint = "2502.07995",
    archivePrefix = "arXiv",
    doi = "10.1103/714c-byxq",
    journal = "Phys. Rev. D",
    volume = "112",
    number = "2",
    pages = "026011",
    year = "2025"
}

@article{Penington:2022dhr,
    author = "Penington, Geoff and Walter, Michael and Witteveen, Freek",
    title = "{Fun with replicas: tripartitions in tensor networks and gravity}",
    eprint = "2211.16045",
    archivePrefix = "arXiv",
    doi = "10.1007/JHEP05(2023)008",
    journal = "JHEP",
    volume = "05",
    pages = "008",
    year = "2023"
}

@article{Gadde:2022cqi,
    author = "Gadde, Abhijit and Krishna, Vineeth and Sharma, Trakshu",
    title = "{New multipartite entanglement measure and its holographic dual}",
    eprint = "2206.09723",
    archivePrefix = "arXiv",
    reportNumber = "TIFR/TH/22-34",
    doi = "10.1103/PhysRevD.106.126001",
    journal = "Phys. Rev. D",
    volume = "106",
    number = "12",
    pages = "126001",
    year = "2022"
}

@article{Jeong:2019xdr,
      author         = "Jeong, Hyun-Sik and Kim, Keun-Young and Nishida,
                        Mitsuhiro",
      title          = "{Reflected Entropy and Entanglement Wedge Cross Section
                        with the First Order Correction}",
      journal        = "JHEP",
      volume         = "12",
      year           = "2019",
      pages          = "170",
      doi            = "10.1007/JHEP12(2019)170",
      eprint         = "1909.02806",
      archivePrefix  = "arXiv",
      SLACcitation   = "%%CITATION = ARXIV:1909.02806;%%"
}

@article{Sohal:2023hst,
    author = "Sohal, Ramanjit and Ryu, Shinsei",
    title = "{Entanglement in tripartitions of topological orders: A diagrammatic approach}",
    eprint = "2301.07763",
    archivePrefix = "arXiv",
    doi = "10.1103/PhysRevB.108.045104",
    journal = "Phys. Rev. B",
    volume = "108",
    number = "4",
    pages = "045104",
    year = "2023"
}

@article{Zou:2020bly,
    author = "Zou, Yijian and Siva, Karthik and Soejima, Tomohiro and Mong, Roger S. K. and Zaletel, Michael P.",
    title = "{Universal tripartite entanglement in one-dimensional many-body systems}",
    eprint = "2011.11864",
    archivePrefix = "arXiv",
    doi = "10.1103/PhysRevLett.126.120501",
    journal = "Phys. Rev. Lett.",
    volume = "126",
    number = "12",
    pages = "120501",
    year = "2021"
}

@article{Berthiere:2023bwn,
    author = "Berthiere, Cl\'ement and Chen, Bin and Chen, Hongjie",
    title = "{Reflected entropy and Markov gap in Lifshitz theories}",
    eprint = "2307.12247",
    archivePrefix = "arXiv",
    doi = "10.1007/JHEP09(2023)160",
    journal = "JHEP",
    volume = "09",
    pages = "160",
    year = "2023"
}

@article{Yamashika:2023qkr,
    author = "Yamashika, Shion and Ares, Filiberto and Calabrese, Pasquale",
    title = "{Time evolution of entanglement entropy after quenches in two-dimensional free fermion systems: A dimensional reduction treatment}",
    eprint = "2310.18160",
    archivePrefix = "arXiv",
    doi = "10.1103/PhysRevB.109.125122",
    journal = "Phys. Rev. B",
    volume = "109",
    number = "12",
    pages = "125122",
    year = "2024"
}

@ARTICLE{2014EL....10740002K,
       author = {{Kormos}, M{\'a}rton and {Bucciantini}, Leda and {Calabrese}, Pasquale},
        title = "{Stationary entropies after a quench from excited states in the Ising chain}",
      journal = {EPL},
         year = 2014,
       volume = {107},
       number = {4},
          eid = {40002},
        pages = {40002},
          doi = {10.1209/0295-5075/107/40002},
archivePrefix = {arXiv},
       eprint = {1406.5070},
       adsurl = {https://ui.adsabs.harvard.edu/abs/2014EL....10740002K}
}

@ARTICLE{2008JSMTE..05..018L,
       author = {{L{\"a}uchli}, Andreas M. and {Kollath}, Corinna},
        title = "{Spreading of correlations and entanglement after a quench in the one-dimensional Bose Hubbard model}",
      journal = {J. Stat. Mech.},
         year = 2008,
       volume = {2008},
       number = {5},
        pages = {05018},
          doi = {10.1088/1742-5468/2008/05/P05018},
archivePrefix = {arXiv},
       eprint = {0803.2947},
       adsurl = {https://ui.adsabs.harvard.edu/abs/2008JSMTE..05..018L}
}

@article{Rieger:2011ipr,
    author = "Rieger, Heiko and Igl{\'o}i, Ferenc",
    title = "{Semi-classical theory for quantum quenches in finite transverse Ising chains}",
    eprint = "1106.5248",
    archivePrefix = "arXiv",
    doi = "10.1103/PhysRevB.84.165117",
    journal = "Phys. Rev. B",
    volume = "84",
    pages = "165117",
    year = "2011"
}

@phdthesis{Lagnese:2022ytv,
    author = "Lagnese, Gianluca",
    title = "{Confinement, Integrability, and Entanglement in Quench Dynamics}",
    school = "SISSA, Italy, SISSA, Trieste",
    year = "2022"
}

@article{Alba:2017lvc,
    author = "Alba, Vincenzo and Calabrese, Pasquale",
    title = "{Entanglement dynamics after quantum quenches in generic integrable systems}",
    eprint = "1712.07529",
    archivePrefix = "arXiv",
    doi = "10.21468/SciPostPhys.4.3.017",
    journal = "SciPost Phys.",
    volume = "4",
    pages = "017",
    year = "2018"
}

@article{Fagotti:2008mlc,
    author = "Fagotti, Maurizio and Calabrese, Pasquale",
    title = "{Evolution of entanglement entropy following a quantum quench: Analytic results for the XY chain in a transverse magnetic field}",
    eprint = "0804.3559",
    archivePrefix = "arXiv",
    doi = "10.1103/PhysRevA.78.010306",
    journal = "Phys. Rev. A",
    volume = "78",
    number = "1",
    pages = "010306",
    year = "2008"
}

@article{Parez:2021pgq,
    author = "Parez, Gilles and Bonsignori, Riccarda and Calabrese, Pasquale",
    title = "{Exact quench dynamics of symmetry resolved entanglement in a free fermion chain}",
    eprint = "2106.13115",
    archivePrefix = "arXiv",
    doi = "10.1088/1742-5468/ac21d7",
    journal = "J. Stat. Mech.",
    volume = "2109",
    pages = "093102",
    year = "2021",
    note = "[Erratum: J.Stat.Mech. 2212, 129901 (2022)]"
}

@article{Alba:2022enm,
    author = "Alba, Vincenzo and Carollo, Federico",
    title = "{Logarithmic negativity in out-of-equilibrium open free-fermion chains: An exactly solvable case}",
    eprint = "2205.02139",
    archivePrefix = "arXiv",
    doi = "10.21468/SciPostPhys.15.3.124",
    journal = "SciPost Phys.",
    volume = "15",
    number = "3",
    pages = "124",
    year = "2023"
}

@article{Parez:2022egh,
    author = "Parez, Gilles and Bonsignori, Riccarda",
    title = "{Analytical results for the entanglement dynamics of disjoint blocks in the XY spin chain}",
    eprint = "2210.03637",
    archivePrefix = "arXiv",
    doi = "10.1088/1751-8121/acb097",
    journal = "J. Phys. A",
    volume = "55",
    number = "50",
    pages = "505005",
    year = "2022"
}

@article{Alba:2017ekd,
    author = "Alba, Vincenzo and Calabrese, Pasquale",
    title = "{Entanglement and thermodynamics after a quantum quench in integrable systems}",
    eprint = "1608.00614",
    archivePrefix = "arXiv",
    doi = "10.1073/pnas.1703516114",
    journal = "Proc. Nat. Acad. Sci.",
    volume = "114",
    number = "30",
    pages = "7947",
    year = "2017"
}

@article{Kim:2013etb,
    author = "Kim, Hyungwon and Huse, David A.",
    title = "{Ballistic Spreading of Entanglement in a Diffusive Nonintegrable System}",
    eprint = "1306.4306",
    archivePrefix = "arXiv",
    doi = "10.1103/PhysRevLett.111.127205",
    journal = "Phys. Rev. Lett.",
    volume = "111",
    number = "12",
    pages = "127205",
    year = "2013"
}

@article{Bertini:2018ymp,
    author = "Bertini, Bruno and Tartaglia, Elena and Calabrese, Pasquale",
    title = "{Entanglement and diagonal entropies after a quench with no pair structure}",
    eprint = "1802.10589",
    archivePrefix = "arXiv",
    doi = "10.1088/1742-5468/aac73f",
    journal = "J. Stat. Mech.",
    volume = "1806",
    number = "6",
    pages = "063104",
    year = "2018"
}

@article{Alba:2017qag,
    author = "Alba, Vincenzo",
    title = "{Entanglement and quantum transport in integrable systems}",
    eprint = "1706.00020",
    archivePrefix = "arXiv",
    doi = "10.1103/PhysRevB.97.245135",
    journal = "Phys. Rev. B",
    volume = "97",
    number = "24",
    pages = "245135",
    year = "2018"
}

@article{DeChiara:2005wb,
    author = "De Chiara, Gabriele and Montangero, Simone and Calabrese, Pasquale and Fazio, Rosario",
    title = "{Entanglement entropy dynamics in Heisenberg chains}",
    eprint = "cond-mat/0512586",
    archivePrefix = "arXiv",
    doi = "10.1088/1742-5468/2006/03/P03001",
    journal = "J. Stat. Mech.",
    volume = "0603",
    pages = "P03001",
    year = "2006"
}

@article{Cotler:2016acd,
    author = "Cotler, Jordan S. and Hertzberg, Mark P. and Mezei, M{\'a}rk and Mueller, Mark T.",
    title = "{Entanglement Growth after a Global Quench in Free Scalar Field Theory}",
    eprint = "1609.00872",
    archivePrefix = "arXiv",
    reportNumber = "MIT-CTP-4827, MIT-CTP-4827",
    doi = "10.1007/JHEP11(2016)166",
    journal = "JHEP",
    volume = "11",
    pages = "166",
    year = "2016"
}

@article{Santalla:2022ygq,
    author = "Santalla, Silvia N. and Ram{\'\i}rez, Giovanni and Roy, Sudipto Singha and Sierra, Germ{\'a}n and Rodr{\'\i}guez-Laguna, Javier",
    title = "{Entanglement links and the quasiparticle picture}",
    eprint = "2208.03766",
    archivePrefix = "arXiv",
    doi = "10.1103/PhysRevB.107.L121114",
    journal = "Phys. Rev. B",
    volume = "107",
    number = "12",
    pages = "L121114",
    year = "2023"
}

@article{Calabrese:2005in,
    author = "Calabrese, Pasquale and Cardy, John L.",
    title = "{Evolution of entanglement entropy in one-dimensional systems}",
    eprint = "cond-mat/0503393",
    archivePrefix = "arXiv",
    doi = "10.1088/1742-5468/2005/04/P04010",
    journal = "J. Stat. Mech.",
    volume = "0504",
    pages = "P04010",
    year = "2005"
}

@article{Caceffo:2024qce,
    author = "Caceffo, Fabio and Alba, Vincenzo",
    title = "{Fate of entanglement in quadratic Markovian dissipative systems}",
    eprint = "2406.15328",
    archivePrefix = "arXiv",
    doi = "10.1103/f6q2-vkyv",
    journal = "Phys. Rev. A",
    volume = "113",
    number = "4",
    pages = "042436",
    year = "2026"
}

@article{Lagnese:2021qqu,
    author = "Lagnese, Gianluca and Calabrese, Pasquale and Piroli, Lorenzo",
    title = "{Entanglement dynamics of thermofield double states in integrable models}",
    eprint = "2112.02008",
    archivePrefix = "arXiv",
    doi = "10.1088/1751-8121/ac646b",
    journal = "J. Phys. A",
    volume = "55",
    number = "21",
    pages = "214003",
    year = "2022"
}

@article{Modak:2020faf,
    author = "Modak, Ranjan and Alba, Vincenzo and Calabrese, Pasquale",
    title = "{Entanglement revivals as a probe of scrambling in finite quantum systems}",
    eprint = "2004.08706",
    archivePrefix = "arXiv",
    doi = "10.1088/1742-5468/aba9d9",
    journal = "J. Stat. Mech.",
    volume = "2008",
    pages = "083110",
    year = "2020"
}

@article{Khetrapal:2024chv,
    author = "Khetrapal, Surbhi and Pedersen, Emil Tore M{\ae}rsk",
    title = "{Mutual information scrambling in an Ising spin chain}",
    eprint = "2402.13558",
    archivePrefix = "arXiv",
    doi = "10.1103/PhysRevB.111.014439",
    journal = "Phys. Rev. B",
    volume = "111",
    number = "1",
    pages = "014439",
    year = "2025"
}

@article{Ghasemi:2021jiy,
    author = "Ghasemi, Mostafa and Naseh, Ali and Pirmoradian, Reza",
    title = "{Odd entanglement entropy and logarithmic negativity for thermofield double states}",
    eprint = "2106.15451",
    archivePrefix = "arXiv",
    doi = "10.1007/JHEP10(2021)128",
    journal = "JHEP",
    volume = "10",
    pages = "128",
    year = "2021"
}

@article{Blass:2012iun,
    author = "Bla{\ss}, Benjamin and Rieger, Heiko and Igl{\'o}i, Ferenc",
    title = "{Quantum relaxation and finite size effects in the XY chain in a transverse field after global quenches}",
    eprint = "1205.3303",
    archivePrefix = "arXiv",
    doi = "10.1209/0295-5075/99/30004",
    journal = "EPL",
    volume = "99",
    pages = "30004",
    year = "2012"
}

@article{Chalas:2024yts,
    author = "Chalas, Konstantinos and Calabrese, Pasquale and Rylands, Colin",
    title = "{Quench dynamics of entanglement from crosscap states}",
    eprint = "2412.04187",
    archivePrefix = "arXiv",
    doi = "10.21468/SciPostPhys.19.5.132",
    journal = "SciPost Phys.",
    volume = "19",
    number = "5",
    pages = "132",
    year = "2025"
}

@article{DiGiulio:2021oal,
    author = "Di Giulio, Giuseppe and Tonni, Erik",
    title = "{Subsystem complexity after a global quantum quench}",
    eprint = "2102.02764",
    archivePrefix = "arXiv",
    doi = "10.1007/JHEP05(2021)022",
    journal = "JHEP",
    volume = "05",
    pages = "022",
    year = "2021"
}

@article{Caceffo:2023hns,
    author = "Caceffo, Fabio and Alba, Vincenzo",
    title = "{Negative tripartite mutual information after quantum quenches in integrable systems}",
    eprint = "2305.10245",
    archivePrefix = "arXiv",
    doi = "10.1103/PhysRevB.108.134434",
    journal = "Phys. Rev. B",
    volume = "108",
    number = "13",
    pages = "134434",
    year = "2023"
}

@article{Ali:2018aon,
    author = "Ali, Tibra and Bhattacharyya, Arpan and Shajidul Haque, S. and Kim, Eugene H. and Moynihan, Nathan",
    title = "{Post-Quench Evolution of Complexity and Entanglement in a Topological System}",
    eprint = "1811.05985",
    archivePrefix = "arXiv",
    reportNumber = "YITP-18-116",
    doi = "10.1016/j.physletb.2020.135919",
    journal = "Phys. Lett. B",
    volume = "811",
    pages = "135919",
    year = "2020"
}

@article{Alba:2018hie,
    author = "Alba, Vincenzo and Calabrese, Pasquale",
    title = "{Quantum information dynamics in multipartite integrable systems}",
    eprint = "1809.09119",
    archivePrefix = "arXiv",
    doi = "10.1209/0295-5075/126/60001",
    journal = "EPL",
    volume = "126",
    number = "6",
    pages = "60001",
    year = "2019"
}

@article{Alba:2019ybw,
    author = "Alba, Vincenzo and Calabrese, Pasquale",
    title = "{Quantum information scrambling after a quantum quench}",
    eprint = "1903.09176",
    archivePrefix = "arXiv",
    doi = "10.1103/PhysRevB.100.115150",
    journal = "Phys. Rev. B",
    volume = "100",
    number = "11",
    pages = "115150",
    year = "2019"
}

@article{Travaglino:2024sxj,
    author = "Travaglino, Riccardo and Rylands, Colin and Calabrese, Pasquale",
    title = "{Quasiparticle picture for entanglement hamiltonians in higher dimensions}",
    eprint = "2412.01538",
    archivePrefix = "arXiv",
    doi = "10.1088/1742-5468/adb7d3",
    journal = "J. Stat. Mech.",
    volume = "2025",
    number = "3",
    pages = "033102",
    year = "2025"
}

@article{Bastianello:2018fvl,
    author = "Bastianello, Alvise and Calabrese, Pasquale",
    title = "{Spreading of entanglement and correlations after a quench with intertwined quasiparticles}",
    eprint = "1807.10176",
    archivePrefix = "arXiv",
    doi = "10.21468/SciPostPhys.5.4.033",
    journal = "SciPost Phys.",
    volume = "5",
    number = "4",
    pages = "033",
    year = "2018"
}

@article{Berthiere:2024sio,
    author = "Berthiere, Cl{\'e}ment",
    title = "{Tripartite entanglement dynamics following a quantum quench}",
    eprint = "2408.12533",
    archivePrefix = "arXiv",
    year = "2024"
}

@article{Bertini_2018,
   title={Entanglement evolution and generalised hydrodynamics: noninteracting systems},
   volume={51},
   ISSN={1751-8121},
   adsurl={http://dx.doi.org/10.1088/1751-8121/aad82e},
   DOI={10.1088/1751-8121/aad82e},
   number={39},
   journal={J. Phys. A},
   publisher={IOP Publishing},
   author={Bertini, Bruno and Fagotti, Maurizio and Piroli, Lorenzo and Calabrese, Pasquale},
   year={2018},
pages={39LT01} }

@article{Calabrese_2020,
   title={Entanglement spreading in non-equilibrium integrable systems},
   ISSN={2590-1990},
   adsurl={http://dx.doi.org/10.21468/SciPostPhysLectNotes.20},
   DOI={10.21468/scipostphyslectnotes.20},
   journal={SciPost Physics Lecture Notes},
   publisher={Stichting SciPost},
   author={Calabrese, Pasquale},
   year={2020} }

@article{Berthiere:2025toi,
    author = "Berthi{\`e}re, Cl{\'e}ment and Gaudin, Paul",
    title = "{Genuine multientropy, dihedral invariants, and Lifshitz theory}",
    eprint = "2509.00593",
    archivePrefix = "arXiv",
    doi = "10.1103/vcqd-rkmn",
    journal = "Phys. Rev. D",
    volume = "113",
    number = "6",
    pages = "065029",
    year = "2026"
}

@article{Iizuka:2025caq,
    author = "Iizuka, Norihiro and Lin, Simon and Nishida, Mitsuhiro",
    title = "{More on genuine multientropy and holography}",
    eprint = "2504.16589",
    archivePrefix = "arXiv",
    doi = "10.1103/x76v-mr6n",
    journal = "Phys. Rev. D",
    volume = "112",
    number = "6",
    pages = "066014",
    year = "2025"
}

@article{Hu:2026bhg,
    author = "Hu, Miao and Lin, Simon and Nechita, Ion",
    title = "{Multi-entropy in random tensor networks}",
    eprint = "2606.04470",
    archivePrefix = "arXiv",
    doi = "10.1007/JHEP08(2026)136",
    journal = "JHEP",
    volume = "08",
    pages = "136",
    year = "2026"
}

@article{Chapman:2018hou,
    author = "Chapman, Shira and Eisert, Jens and Hackl, Lucas and Heller, Michal P. and Jefferson, Ro and Marrochio, Hugo and Myers, Robert C.",
    title = "{Complexity and entanglement for thermofield double states}",
    eprint = "1810.05151",
    archivePrefix = "arXiv",
    primaryClass = "hep-th",
    doi = "10.21468/SciPostPhys.6.3.034",
    journal = "SciPost Phys.",
    volume = "6",
    number = "3",
    pages = "034",
    year = "2019"
}

\let\addcontentsline\oldaddcontentsline

\end{document}